\documentclass[twocolumn]{aastex631}

\newcommand{\COO}{CO$_2$ }
\newcommand{\OOO}{O$_3$ }
\newcommand{\cmu}{cm$^{-1}$ }

\usepackage{chemfig}
\usepackage[version=4]{mhchem}

\usepackage[dvipsnames]{xcolor}

\begin{document}

\title{Functionalization of Benzene Ices by Atomic Oxygen}

\author[0000-0001-6947-7411]{Elettra L. Piacentino}
\affiliation{Harvard-Smithsonian Center for Astrophysics, 60 Garden Street, Cambridge, MA 02138, USA }

\author[0000-0002-2295-5452]{Alexandra McKinnon}
\affiliation{Department of Chemistry and Chemical Biology, Harvard University, 12 Oxford Street, Cambridge, MA 02138, USA}

\author[0000-0002-8253-6436]{Nora H\"anni}
\affiliation{Space Research $\&$ Planetary Sciences,Physics Institute, University of Bern, Sidlerstrasse 5, 3012 Bern, Switzerland.}

\author[0000-0002-3258-3509]{Amit Daniely}
\affiliation{Fritz Haber Research Center for Molecular Dynamics, The Hebrew University of Jerusalem, Jerusalem 9190401, Israel}

\author[0000-0003-0701-9997 ]{Estefania Rossich Molina}
\affiliation{Fritz Haber Research Center for Molecular Dynamics, The Hebrew University of Jerusalem, Jerusalem 9190401, Israel}

\author[0000-0003-3155-2273]{Tamar Stein}
\affiliation{Fritz Haber Research Center for Molecular Dynamics, The Hebrew University of Jerusalem, Jerusalem 9190401, Israel}

\author[0000-0002-8716-0482]{Jennifer Bergner}
\affiliation{UC Berkeley, Department of Chemistry, Berkeley, California 94720, USA}

\author[0000-0003-2761-4312]{Mahesh Rajappan}
\affiliation{Harvard-Smithsonian Center for Astrophysics, 60 Garden Street, Cambridge, MA 02138, USA }

\author[0000-0001-8798-1347]{Karin I. {\"O}berg}
\affiliation{Harvard-Smithsonian Center for Astrophysics, 60 Garden Street, Cambridge, MA 02138, USA }

\begin{abstract}
Small aromatic molecules, including functionalized derivatives of benzene, are known to be present throughout the different stages of star and planet formation. In particular, oxygen-bearing monosubstituted aromatics, likely including phenol, have been identified in the coma of comet 67P. This suggests that, earlier in the star and planet formation evolution, icy grains may act as both reservoirs and sites of functionalization for these small aromatics. We investigate the ice-phase reactivity of singlet oxygen atoms (O($^1$D)) with benzene, using ozone as a precursor that is readily photodissociated by relatively low-energy. Our experiments show that O($^1$D) efficiently reacts with benzene, forming phenol, benzene oxide, and oxepine as the main products.  
Phenol formation is temperature-independent, consistent with a barrierless insertion mechanism. In contrast, the formation of benzene oxide/oxepine shows a slight temperature dependence, suggesting that additional reaction pathways involving either ground-state or excited-state oxygen atoms may contribute. In H$_2$O and \COO ice matrices we find that dilution does not suppress formation of phenol. 
We extrapolate an experimental upper limit for the benzene-to-phenol conversion fraction of 27–44$\%$ during the lifetime of an interstellar cloud, assuming O($^1$D) production rates based on CO$_2$ ice abundances and a cosmic-ray induced UV field.
We compare these estimates with a new analysis of data from the comet 67P, where the C$_6$H$_6$O/C$_6$H$_6$ ratio is 20±6$\%$.
This value lies within our estimated range, suggesting that O($^1$D)-mediated chemistry is a viable pathway for producing oxygenated aromatics in cold astrophysical ices,  
potentially enriching icy planetesimals with phenol and other biorelevant compounds.
\end{abstract}

\section{Introduction} \label{sec:intro}

Aromatic molecules are essential to biology, forming the structural core of deoxyribonucleic acid (DNA) and ribonucleic acid (RNA) bases, several amino acids, and key metabolic coenzymes (see, e.g., \citet{szatylowicz2021aromaticity, balaban2004aromaticity} and \citet{pozharskii2011heterocycles}). Benzene, as the prototypical aromatic molecule, serves as a foundational model for understanding the behavior of this important class of molecules.

% Prior to 2018, interstellar aromatic molecules were primarily detected in the form of large polycyclic aromatic hydrocarbons (PAHs) 
Prior to 2018, interstellar detection of aromatic functionalities was limited to families of large polycyclic aromatic hydrocarbons (PAHs)
\citep[e.g.,][]{leger1984identification,allamandola1989interstellar,tielens1997circumstellar,tielens2013molecular,gudipati2003ApJ...596L.195G,gudipati2006ApJ...638..286G,bouwman2010A&A...511A..33B,cook2015photochemistry,Zeichner2023Sci...382.1411Z}.
% . PAHs have been known for many decades to exist in a variety of astronomical objects, but 
However the detection of smaller aromatic molecules has become more frequent in recent years. Benzene was tentatively detected for the first time in the protoplanetary nebula CRL 618 by \citet{cernicharo2001infrared}. Since then many small benzene derivatives, such as benzonitrile \citep{mcguire2018Sci...359..202M}, have been detected in star forming region of the interstellar medium (ISM). Recently, benzene has also been detected in the inner disk of the J160532 M dwarf star \citep{tabone2023rich} as well as in the disk around the very low mass star ISO-ChaI 147 \citep{arabhavi2024abundant}. In comets, the aromatic inventory also appears to be significant. Benzene and toluene were detected in the coma of comet 67P during the Rosetta mission's May 2015 flyby \citep{schuhmann2019aliphatic}. Additional oxygen- and nitrogen-bearing aromatic compounds were identified in the data from 3 August 2015.
% flyby. 
In that dataset, benzaldehyde and benzoic acid were detected with high confidence \citep{hanni2022identification}, 
% \citep{hanni2023oxygen}
while the identifications of phenol \citep{hanni2023oxygen} and benzonitrile \citep{Hanni2025nitrogenA&A...699A.135H} were less certain due to degeneracies/ambiguities in the deconvolution of the complex mass spectrum.
% ambiguous fragmentation patterns. 
The variety of functionalized aromatic molecules detected indicates that star and planet formation are characterized by a rich aromatic astrochemistry, suggesting that several reaction pathways must be available to convert simple parent molecules like benzene into more complex, functionalized compounds.

Several experimental and theoretical studies have explored gas-phase pathways leading to the functionalization of benzene and other unsubstituted aromatics. For example, benzonitrile can be efficiently formed in the gas-phase through the direct addition of a cyano group to a benzene molecule \citep{cooke2020benzonitrile}. Theoretically, it has also been shown that benzonitrile can form upon ionization of trimer clusters containing one cyanoacetylene group \citep{jose2021molecular}. Gas-phase reactions involving oxygen atoms have also been investigated. In particular, the reaction between pyridine and excited-state oxygen atoms has been shown to proceed at low temperatures, resulting in the formation of oxygenated addition products \citep{Recio22}.

While gas phase chemistry is important, molecules such as benzene and its derivatives tend to efficiently freeze out and then remain on the ice even at temperatures higher than water desorption \citep{Piacentino2024ApJ...974..313P}. As a result, the ice-phase chemistry of this class of molecules is expected to be most significant
% chemistry of this class of molecules is expected to be mostly occurring in the ice phase 
throughout the different stages of star and planet evolution. Ice irradiation experiments indicate that exposure to vacuum-ultraviolet photons (VUV), or electron irradiation, triggers radical-radical reactions in benzene ices. For instance, carboxylic acids form when benzene or pyridine ices mixed with \COO are irradiated at 10 K \citep{mcmurtry2016formation}, while benzonitrile forms upon electron exposure of ices composed of benzene and acetonitrile \citep{maksyutenko2022formation}. Heterocycles can also form from UV-driven processes in ices. As demonstrated by \citet{materese2015n}, for UV irradiated mixtures of benzene or naphthalene embedded in ices containing H$_2$O and NH$_3$.

So far the role of excited-state reactants in mediating ice-phase functionalization of aromatics has not been explicitly considered under conditions relevant to astrophysical environments. Singlet oxygen atoms (O($^1$D)) are easily formed from the photolysis of molecules that are very abundant in interstellar ices. Photons in the 100-200 nm range can trigger the photolysis of O$_2$, CO$_2$, and water to produce O($^1$D) with high efficiency. 
In particular, the formation efficiency for O($^1$D) is close to unity in both the dissociation of molecular oxygen \citep{lee1977quantum} and \COO \citep{Okabe78,Slanger1982JChPh..77.2432S}, and is approximately 10-23\,$\%$ for water dissociation at 120 and 147\,nm, respectively
% 10\,$\%$ for water
\citep{Slanger1982JChPh..77.2432S,Ung1974CPL....28..603U}. Based on experiments with aliphatic hydrocarbon ices, excited state oxygen atoms can insert into or add to the C-H and C-C bonds leading to the formation of  alcohols, aldehydes, and oxides, even at the very low temperatures typical of interstellar medium (ISM) ices \citep{bergner2017methanol,bergner2019detection}.
The condensed-phase reaction of benzene with oxygen atoms was initially explored at cryogenic temperature by \citet{parker1999photochemical}, who irradiated benzene in argon matrices at wavelengths above 280 nm and identified phenol as the major reaction product but also identified benzene oxide as a secondary isomer. 

Building on the previous work, we investigate the ice-phase chemistry of benzene with photoproduced singlet state oxygen in model ices costituted of solely benzene and the laboratory O($^1$D) precursor, ozone, as well as in ice mixtures that include \COO or H$_2$O.
Through our experiments, we provide a detailed characterization of the reaction products under fiducial conditions and extend the study to various temperatures, ice ratios, and additional matrix components to evaluate these parameters and better extrapolate their effects on realistic ISM ices.
We present a complete description and product characterization of the fiducial experiment in Sections \S\ref{subsub:reaction products fid} and \S\ref{sub:growth curves}. In Sections \S\ref{res:ratio}, \S\ref{res:temp} and \S\ref{res:mixture}, we describe the influence of ice ratio, temperature and ice composition on the chemistry and we discuss it in \S\ref{sec:disc}. Finally, \S \ref{astrochem}  discusses the astrochemical relevance of our findings.

\section{Experimental Details}\label{methods}

\subsection{Experimental set-up}
The experiments were run using the SPACECAT (Surface Processing Apparatus for Chemical Experimentation to Constrain Astrophysical Theories) set-up which has been described in detail elsewhere \citep{martin2020formation}. In brief, SPACECAT consists of a Ultra-High Vacuum (UHV) spherical chamber that is kept at a pressure of 10$^{-10}$ torr. The gas phase mixtures are delivered to the chamber via a leak valve assembly and condensed onto a helium-cooled CsI substrate that is placed in the center of the SPACECAT chamber. The ice constituents can be delivered through separate dosers, or premixed before deposition. 
The temperature of the substrate is controlled between 12 and 300 K ($\pm$2 K accuracy and uncertainty of 0.1 K) using a LakeShore 335 temperature controller. The ice coverage and composition are monitored using a Fourier transform infrared spectrometer (Bruker Vertex 70v), while the composition of the gas can be monitored using a Pfeiffer QMG 220M1 mass spectrometer directly connected to the chamber.
SPACECAT is also equipped with a port dedicated to photon processing to which a Analytik Jena UVP Mercury Pen-Ray Lamp (Model 11SC-1) is attached. The emission band of the lamp is focused at 254\,nm and produces a photon flux of $\sim$\,10$^{13}$ photons per cm$^2$ per second. The photon flux is measured at the beginning of each experiment using a National Institute of Standards and Technology (NIST) calibrated AXUV-100G photodiode that has an uncertainty of $\sim$5$\%$ \citep{bergner2019detection}.

\subsection{The formation of \texorpdfstring{O($^1$D)}{O(1D)}}
\label{methods:subse-ozoneprod}

Ozone is a useful source of oxygen atoms in laboratory interstellar ices analogues \citep{brann2025methyl}. Unlike other sources of oxygen atoms commonly used in ice experiments, such as CO$_2$ and O$_2$ \citep{bergner2017methanol,bergner2019oxygen}, which require wavelengths below 160 nm to produce O($^1$D), ozone can be dissociated at a lower energy of 254 nm. This has the advantage that many organic molecules do not fragment at this wavelength, allowing us to isolate the reaction triggered by atomic oxygen species in the experiment without interference from organic radicals.

In our experiments, ozone is produced external to the vacuum chamber from molecular oxygen before each experiment using a Nano 15 Ozone Generator (Absolute Ozone) \citep{brann2025methyl}.  A continuous flow of O$_2$ is passed through the generator where an electric discharge produces \OOO directly from the high purity reagent. The flowing ozone is then directly condensed onto the cooled substrate via a dedicated leak valve with a deposition rate of $\sim$\,3\, monolayers (ML) per min. To avoid ice contamination with leftover molecular oxygen we prepare the ices at 40\,K, a temperature above the sublimation temperature of O$_2$. After deposition we cool the ice to the desired experimental temperature.

The photodissociation of ozone in the ice strongly depends on the wavelength of incident photons, which determines the electronic state of the oxygen atoms produced. Photons within the strong Hartley absorption band ($\sim$\,225–305\,nm) predominantly yield singlet oxygen atoms, O($^1$D), whereas wavelengths above 400 nm selectively generate triplet ground-state oxygen atoms, O($^3$P) \citep{matsumi2003photolysis}. In our experiments, O($^1$D) is produced directly in the ice phase by irradiating \OOO with 254\,nm photons, which corresponds to the maximum of the Hartley absorption band. At this wavelength, the reaction, \OOO + h$\nu$ $\rightarrow$ O($^1$D) + O$_2$(a$^1\Delta$g), has been measured to proceeds with a quantum yield of 0.92 \citep{takahashi2002quantum}. 

Once produced, O($^1$D) can survive in the ice phase before relaxing to the ground atomic state O($^3$P). The lifetime of metastable O($^1$D$)$ has been found to be matrix-dependent, lasting for a few hundred milliseconds in SF$_6$ matrices \citep{mohammed19901} and approximately 32 s in neon \citep{Fournier1982CP.....70...39F}. 
% Once produced, O($^1$D) can survive in the ice phase for up to a few hundred milliseconds before relaxing to the ground atomic state O($^3$P) \citep{ennis2012formation}. 
Given the relatively long lifetime and the high quantum yield associated with its formation, we can reasonably assume that the vast majority of atomic oxygen generated under our experimental conditions is the singlet state O($^1$D). The other major photo fragment, O$_2$(a$^1\Delta$g), exhibits a much shorter lifetime that is strongly dependent on the surrounding environment. For example, in benzene solutions, its lifetime has been measured to be $\sim$\,26.7 $\mu$s \citep{ogilby1982chemistry}. Given its significantly shorter lifetime, we consider it unlikely that O$_2$(a$^1\Delta$g) undergoes substantial chemistry under our experimental conditions. Taken together, these considerations support the conclusion that, at the fiducial experimental temperature, nearly all reactive oxygen atoms generated in our system is present as singlet atomic oxygen, O($^1$D).

Although quenching collisions with undissociated ozone molecules can lead to the loss of a fraction of O($^1$D) \citep{matsumi2003photolysis,giachardi1972photolysis}, this loss is substantially reduced in our system by employing a benzene-dominated ice. However, in diluted ices, quenching of O($^1$D) can also occur through collisions with water and CO$_2$ molecules. In the case of water in particular, this process may be significant, as O($^1$D) reacts efficiently with H$_2$O to produce OH radicals \citep{vranckx2010kinetics}. These radicals can subsequently recombine to form H$_2$O$_2$ \citep{zheng2007mechanistical} and generate O($^3$P) atoms \citep{hama2009formation}, resulting in a greater variety of reactive species present in the ice.

\subsection{Reagents and ice characterization}

Samples are prepared using off the shelf reagents having the following specifications:
C$_6$H$_6$ (Millipore sigma, 99.8$\%$), oxygen (Airgas, 99.99 atom $\%$ $^{16}$O), CO (Millipore sigma, 99.9$\%$), CO$_2$ (Millipore sigma, 99.9$\%$), and purified water. Gas-phase reagents were used directly without transfers or purification. Condensed-phase reagents were transferred into resealable glass flasks and purified using several freeze-thaw-pump cycles using liquid nitrogen.

The initial ice coverage for each infrared-active molecule in the ice are calculated from the IR spectra using Eq. \ref{eq:ml} 

\begin{equation} \label{eq:ml}
\rm N = \frac{2.3}{\rm A} \int Abs(\tilde\nu) d\tilde\nu
\end{equation}

where, N (molecules cm$^{-2}$) is the column density of the molecule, which can be expressed as coverage in monolayers (ML), assuming 1 ML $\approx$ 10$^{15}$ molecules cm$^{-2}$. A (cm molecule$^{-1}$) is the band strength, and $\int$Abs($\tilde\nu$) d$\tilde\nu$ is the integrated area of the infrared (IR) band in absorbance units. The band strengths used in this work are reported in Table~\ref{table:IR_BS}.

\begin{deluxetable}{c||ccc}[t]
\tablecolumns{4}
\tablewidth{0.8\columnwidth}
\label{table:IR_BS}
\tablecaption{Band strengths values.}
\tablehead{
Molecule & Band position & A/10$^{-18}$ 10 K         & Ref.\\
         & (\cmu)     & (cm molecule$^{-1}$)      &     }
\startdata
O$_3$            & 1035   &  8.8 &  $^{[1]}$ \\
C$_6$H$_6$       & 1477	& 4.8 &  $^{[2]}$ \\
C$_6$H$_5$OH     & 1503	& 7.8 &  $^{[3]}$$^{\dagger}$ \\
H$_2$O         & 3280 	& 200 &  $^{[4]}$ \\
CO$_2$         & 2340   & 110  &  $^{[5]}$ \\
\enddata
\tablenotetext{}{$^{[1]}$\citet{Loeffler06}, $^{[2]}$\citet{hudson2022infrared}, $^{[3]}$\citet{Piacentino2024ApJ...974..313P}, $^{[4]}$\citet{Gerakines95}, $^{[5]}$\citet{Bouilloud2015MNRAS.451.2145B}.}
$^{\dagger}$ The band strength of phenol ice from \citet{Piacentino2024ApJ...974..313P} was derived by comparison with that of CO. Due to the very different volatilities, the band strength of phenol carries an uncertainty of $\sim$\,50 $\%$.   
\end{deluxetable}

\subsection{Ice Destruction and Formation Kinetic Modeling}\label{methods:subsec-prodquantification}

To quantify both reactant loss and product formation during irradiation, we integrate the IR absorption bands of the relevant species as a function of photon fluence. We then apply kinetic models based on first order behavior with respect to photon flux, under the steady-state approximation.

To model the destruction of reactants, we follow the first order kinetic model under the steady-state approximation described by Piacentino (2025, under review), and given in Eq. \ref{eq:decay},

\begin{equation} \label{eq:decay}
   \rm N_{Rg}(\phi)/N_{Rg}(\phi=0) = e^{-\sigma_{Ice}\phi} +C
\end{equation}

where N$_{Rg}(\phi)/N_{Rg}(\phi$=0) indicates the rate of disappearance of the reagent molecule (Rg) from the ice as a function of photon fluence $\phi$ (photons/cm$^{-2}$). While the factor C is the steady-state abundance that accounts for the balance between all the processes including destruction, recombination and other chemical reactions that are occurring in the ices. Finally, $\sigma_{Ice}$ (cm$^2$) represents the total cross section for the disappearance of the reagent molecules from the ice, and is a linear combination of the cross sections for all the processes occurring within the ice.

The IR bands corresponding to reaction products are also analyzed as a function of photon fluence to determine both the formation cross section and the steady-state yield for each observed product. This approach follows the methodology of \citet{bergner2019oxygen}, by evaluating the integrated area of the bands during the irradiation phase of the experiment and modeling the growth curve using a first order kinetic law derived under the steady-state approximation (Eq. \ref{eq. integrated law}) 

\begin{equation} \label{eq. integrated law}
    \rm N_{Pr}(\phi)/N_{Pr} (\phi=\infty) =\left( 1 - e^{-\sigma_{Form} \phi} \right)
\end{equation}

Where N$_{Pr}(\phi)/N_{Pr} (\phi=\infty$) is the rate of formation of the products (Pr), N$_{Pr}(\phi=\infty$) is the steady state column density, $\phi$ the photon fluence, and $\sigma$$_{Form}$ represent the observed net formation cross sections, which is again a linear combination of formation and destruction reactions.

\subsection{Uncertainty on the kinetic constants} \label{sec:uncertainty}

For the destruction cross sections, we adopt an uncertainty of 11\,$\%$, consistent with the value reported by Piacentino (2025, under review). This value accounts for experimental variability, uncertainties from data fitting, and variations in the photon flux to which the ices are exposed. Uncertainties in the IR band strengths are implicitly accounted for in our experiments because the destruction cross sections are measured relative to the initial ice abundance and are therefore unaffected by this source of error, which only influence the estimation of ice coverage. Including the same sources of error used for the destruction cross sections, we also report the uncertainty on the steady state abundances to be \,20\,$\%$.
The uncertainty in the formation cross sections ($\sigma_\mathrm{Form}$) is calculated by combining in quadrature three sources: variability across three repeats of the fiducial experiments (Exp. 1, 2, and 3 in Table \ref{tab_explist}), the fitting error on the growth curves, and a 5\,$\%$ uncertainty on the lamp flux calibration. Band strength errors are not included, as formation cross sections are calculated from values normalized to the final abundance of each species.  A detailed treatment of the uncertainty is presented in App. \ref{app:uncertainty rep}. On average the formation cross section evaluation of uncertainty yields $\sim$ 23 $\%$ for all the bands in Fig \ref{fig:fit fiducial} with the exception of the bands at 1073 and 760 \cmu for which the uncertainty on the formation cross section is up to $\sim$ 34 $\%$ due to more variability across experiments and lower signal to noise.

\begin{figure*} [t]
\centering
\includegraphics[width=\textwidth]{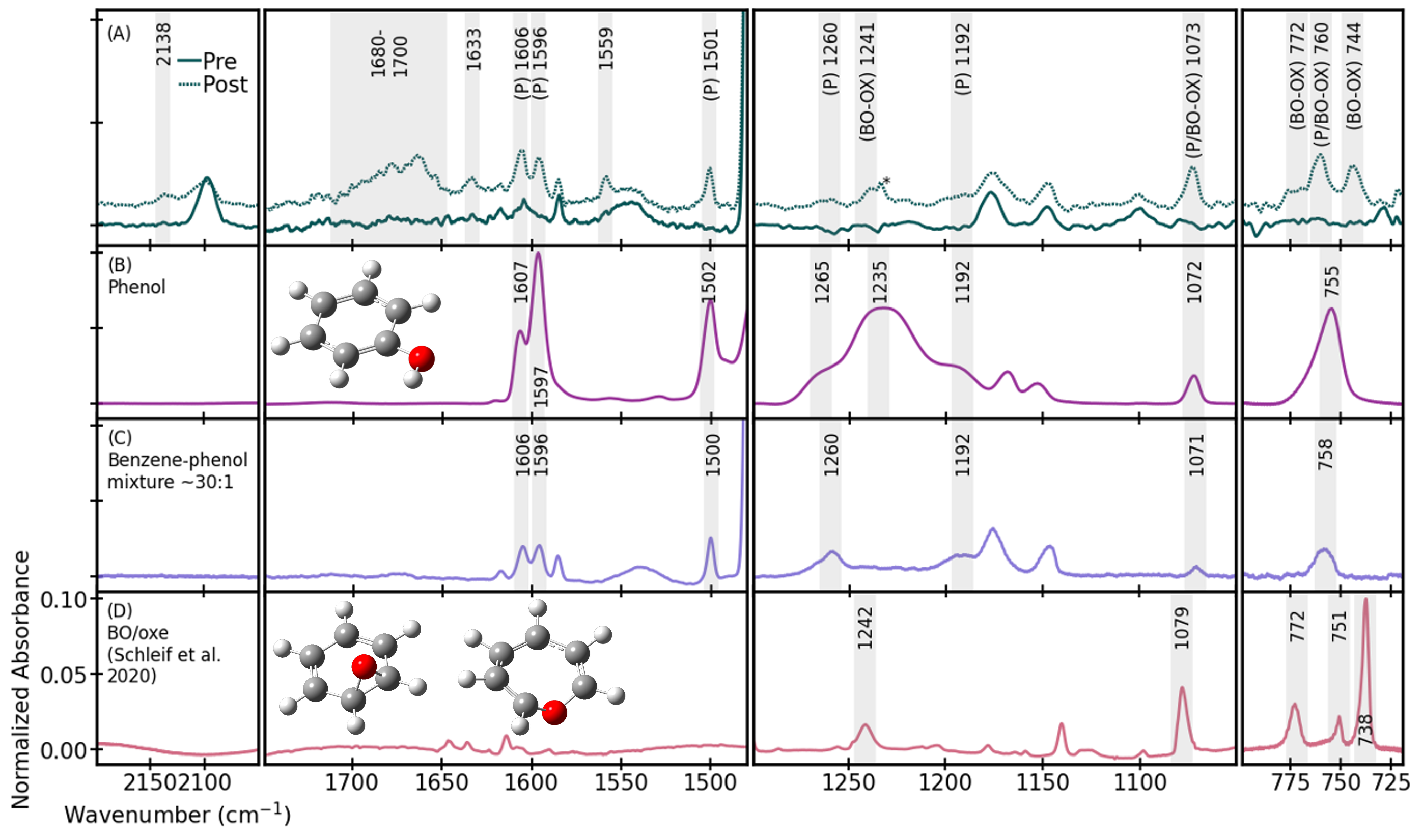}
\caption{(A) Pre and post irradiation spectra of the fiducial 10 K 2:1 benzene:ozone experiment. (B) Spectra of pure phenol ice at 10 K, (C) Spectra of a $\sim$ 30:1 benzene:phenol mixture at 10 K (D) Spectra of a benzene-oxide/oxepine mixture at 10 K reproduced with permission from \citet{Schleif2020AngCh}. Optimized geometries representative of phenol, benzene oxide and oxepine are obtained using Gaussian 16 suite of software and are shown for reference in the corresponding panel. For clarity, only relevant spectral region are shown, full spectral data are available for download at 10.5281/zenodo.17216222 and 10.5281/zenodo.17216256}
\label{fig:spectralmatch}
\end{figure*}

\section{Experimental Results}\label{sec:results}

We first present the result of the fiducial experiment, which is a 2:1 mixture of benzene:ozone irradiated at 10\,K with 254\,nm photons, which we use to identify the major products (\S\ref{subsub:reaction products fid}), and to evaluate the reagent destruction and product formation cross sections (\S\ref{sub:growth curves}). 
We then study the effect of varying ice ratio (\S\ref{res:ratio}), temperature (\S\ref{res:temp}) and composition (\S\ref{res:mixture}) on both the reaction products and on the kinetics.

\begin{deluxetable*}{cc||cccccccccc}[t]
\tablecolumns{7}
\tablewidth{\textwidth}
\label{tab_explist}
\tablecaption{List of experiments and main kinetic parameters derived for each experiment.}
\tablehead{&Mixture &  Ratio& C$_6$H$_6$  &  T &   Max.  & \multicolumn{2}{c}{C$_6$H$_6$}& \multicolumn{2}{c}{O$_3$} \\
&& & Cov.& & Fluence & $\sigma$$_{Ice}$ &  1-C  &$\sigma$$_{Ice}$ & 1-C  \\
\hline
&& & (ML)& (K)& (cm$^{-2}$/10$^{16}$) & (cm$^{2}$/10$^{-17}$)&($\%$) &(cm$^{2}$/10$^{-17}$)&($\%$)}
\startdata 
1 &C$_{6}$H$_{6}$:O$_{3}$         & 2:1  & 27  & 10 & 9.1& 1.5 &26 & 1.3 &81\\ %27b(21) 15-17o(8) 110824 1:1.6-1.8 1.08-002 fid chosing the multiple gaussian fit for all
2 &C$_{6}$H$_{6}$:O$_{3}$         & 2:1  & 26  & 10 & 8.9& 1.5 & 21 &1.4& 76 \\ %26b 11-13o 102524 1:2.4-2 1.06-0.02
3 &C$_{6}$H$_{6}$:O$_{3}$         & 1:1  & 15  & 10 & 9.5& 1.4 &36 &1.2 &77\\ %26b 11-13o 20250424 1:2.4-2 1.13-0.01
\hline
4 &C$_{6}$H$_{6}$:O$_{3}$         & 4:1  & 42  & 10 & 9.0& 1.3 &13 & 1.6 &82 \\ 
5 &C$_{6}$H$_{6}$:O$_{3}$         & 10:1  & 59  & 10 & 9.1& 0.5 & 6 & 1.4 &80 \\ 
\hline
6 &C$_{6}$H$_{6}$:O$_{3}$         & 2.5:1  & 20  & 20 & 10.0 & 1.2 &17 & 1.2 &68 \\ %20b 8o 042925 1:2.2 1.175-0.01
7 &C$_{6}$H$_{6}$:O$_{3}$         & 2:1  & 20  & 40 & 9.4& 1.5 &18 & 1.3 &86 \\ %20b 9o 051124 1:2.2 1.11-0.017, 
\hline
8 &C$_{6}$H$_{6}$:O$_{3}$:\COO        & 1:0.75:4   & 16 & 10 & 9.0& 1.4 &40 & 1.1 &76 \\%16:12:65 ozone by difference in the area \\ 200250425 1.174-0.009
9 &C$_{6}$H$_{6}$:O$_{3}$:H$_2$O      & 1:0.5:4.8          & 13 & 10 & 9.2& 1.5 &20 & 1.7 &74\\ % 13:6:63 * ozone estimated with questionable three gaussian integartion, subtraction metho does not work  1.24-0.09
\hline
10 & C$_{6}$H$_{6}$                 & -    & 17  & 10 &9.1&-&-&-&-\\ %no lamp recorderd assumed 1.08-0.02
11 & C$_{6}$H$_{6}$:Xe              & 1:3* & 21  & 10 &9.0&-&-&-&-\\ %no lamp recorderd assumed 1.06-0.02
12 & C$_{6}$H$_{6}$:C$_{6}$H$_{5}$OH & 35:1 & 104 & 10 & -&-&-&-&-\\
13 &C$_{6}$H$_{5}$OH:O$_{3}$        &   1:2.5        & 10 \textdagger & 10 & 9.2&-&-&-&-\\ %no data on the lamp current estimated 1.08-0.02 21225
\hline
\enddata
\tablecomments{ $^*$ Gas phase ratio, \textdagger Phenol ice coverage. The experiments are grouped by purpose: Exp. 1 serves as the fiducial case. Exp. 2–3 are repeats used to evaluate experimental uncertainty. Exp. 4–5 explore variations in the ice composition ratio. Exp. 6–7 investigate the effects of different ice temperatures. Exp. 8–9 are mixtures with H$_2$O or CO$_2$. Exp. 10–13 are additional diagnostic tests.}
\end{deluxetable*}

\subsection{Reaction Products in the fiducial experiment}\label{subsub:reaction products fid}

Based on previous studies on hydrocarbons \citep{bergner2017methanol, bergner2019detection, daniely2025photochemical}, we expect oxygen atoms to react barrierlessly with benzene via insertion into C–H bonds, yielding phenol (C$_6$H$_5$OH) as the primary alcohol product.
The spectral variation observed in the fiducial experiment is presented in panel A of Fig.\ref{fig:spectralmatch}, alongside the IR ice spectra of the primary reaction products, phenol (Panel B), benzene oxide, and oxepine (Panel D). The phenol IR spectrum was measured at 10 K using SPACECAT, while the benzene oxide/oxepine mixture spectrum, measured in an argon matrix at 4 K, was reproduced with permission from \citet{Schleif2020AngCh}.  
Product band assignment is shown in Fig.\ref{fig:spectralmatch}, however a summary of the IR bands formed during irradiation and their spectral assignments is also provided in Table \ref{tab:productsoffiducial}.

Comparison of the fiducial experiment spectra and the one of pure phenol show substantial similarities (Fig. \ref{fig:spectralmatch}). Both the bands at $\sim$\,1600 and 1500 \cmu, corresponding to ring stretching modes of phenol, are visible in the fiducial experiment post irradiation spectra. The hydrogen out-of-plane mode resonating at 755 \cmu in the pure phenol ice, is observed at 760 cm$^{-1}$ in the fiducial experiment, exhibiting a minor shift of 5 cm$^{-1}$. Finally, the combination bands in the 1200 \cmu region of the phenol spectra are also apparent in the fiducial experiment spectra after irradiation. While the similarities are notable, several differences in band shape are also evident. Specifically, we find that the relative ratio in the individual peaks within the bands at 1600 \cmu and in the 1192-1265 \cmu in the pure phenol spectrum do not match those observed in the post-irradiation spectra of the fiducial experiment. Furthermore, the band shifts at lower wavenumbers are also significant.

To better understand the observed spectral differences, we recorded the IR spectra of a benzene:phenol mixture at an approximate ratio of 30:1 (Fig. \ref{fig:spectralmatch}, Panel C). At this ice dilution, the intermolecular interactions between adjacent phenol molecules are reduced resulting in notable spectral changes that more closely resemble the results of the fiducial experiment. Notably, the relative intensity of the ring stretching mode near 1600 cm$^{-1}$ in the benzene-isolated phenol spectrum (Fig. \ref{fig:spectralmatch} Panel C) better matches the corresponding peak ratio observed in the fiducial data. Additionally, the strong and broad band at 1235 cm$^{-1}$, present in the pure phenol spectrum, is absent in the benzene-isolated mixture, aligning more closely with the fiducial post-irradiation spectra. The hydrogen out-of-plane bending mode also shifts slightly, from 755 cm$^{-1}$ in pure phenol to 758 cm$^{-1}$ in the benzene mixture, allowing for a closer match to the 760 cm$^{-1}$ feature observed in the fiducial spectrum. Finally, the distinct band at 1073 cm$^{-1}$ in the fiducial experiment could plausibly be attributed to the 1072 cm$^{-1}$ mode of phenol. However its relatively high intensity in the fiducial experiment compared to other characteristic phenol bands does not match neither the pure phenol nor the phenol-benzene ice spectra, suggesting that phenol may not be the only species contributing to this absorption. This discrepancy points to the possible presence of additional reaction products having overlapping vibrational modes from other molecules within the ice matrix.

Alongside inserting into the C–H bonds of benzene, oxygen atoms are also expected to react with the C–C multiple bonds via an addition process, producing benzene oxide. This reaction is not expected to have an activation barrier for O($^1$D). However, previous studies have shown that analogous addition reactions, sometimes mediated by ground state oxygen atoms \citep{vanuzzo2021crossed}, such as the formation of c-C$_2$H$_4$O from C$_2$H$_4$, can still exhibit temperature dependence \citep{bergner2019detection}.  The comparison of the fiducial spectrum  with that of the benzene oxide/oxepine mixture also reveals several matching features (Fig. \ref{fig:spectralmatch}, Panel D). In particular, in the 700–800 cm$^{-1}$ region, the bands at 738 and 772 cm$^{-1}$ in the benzene oxide/oxepine spectrum align well with the 743 cm$^{-1}$ band and the 772 cm$^{-1}$ shoulder observed in the fiducial data. The smaller benzene oxide/oxepine band at 751 \cmu is also likely to contribute to the 760 \cmu band formed in the fiducial experiment. At slightly higher wavenumbers, the band at 1241 cm$^{-1}$ in the fiducial spectrum can be assigned to the 1242 \cmu feature in the benzene oxide/oxepine spectra, although its assignment remains uncertain due to partial overlap with a known spectral artifact. Finally, the 1079 cm$^{-1}$ band in the benzene oxide/oxepine reference spectrum may contribute to the 1073 cm$^{-1}$ feature in the fiducial experiment.

At higher wavenumbers, several IR features cannot be directly assigned to either phenol or the benzene oxide/oxepine. However, two bands may be tentatively attributed to secondary products formed from the reaction of phenol with singlet oxygen. To investigate this, we performed a dedicated phenol:ozone experiment under similar conditions to the fiducial (App. \ref{App: secondary}, Fig. \ref{figapp: secondary product fit}). The resulting spectrum includes a weak feature at 2138\,cm$^{-1}$ and a broad band in the 1650–1700\,cm$^{-1}$ region, both of which closely resemble bands observed in the fiducial experiment. We tentatively assign the 2138\,cm$^{-1}$ feature to the R–C=C=O stretch of a ketene formed via ring opening, as previously proposed by \citet{parker1999photochemical}.
Finally, the two sharp bands at 1559 and 1633\,cm$^{-1}$, evident in the fiducial spectra, do not appear to correspond directly to any of the expected primary or secondary reaction products and remain unassigned.

\begin{deluxetable*}{ccc||ccc||ccc}[t]
\tablecolumns{7}
\tablewidth{\textwidth}
\label{tab:productsoffiducial}
\tablecaption{Product bands and assignment for the mixed ices experiment}
\tablehead{\multicolumn{3}{c}{C$_6$H$_6$:O$_3$} &\multicolumn{3}{c}{C$_6$H$_6$:O$_3$:\COO} & \multicolumn{3}{c}{C$_6$H$_6$:O$_3$:H$_2$O}\\ \hline Band (\cmu) & ID & Ref. & Band (\cmu) & ID & Ref.& Band (\cmu) & ID & Ref.} 
\startdata 
743  & BO-OX   & [1,\,2]  & 747  & BO-OX   & [1,\,2] &   -  &   -    &  -\\%& 744  &   P,\,BO  & [1,\,2]  \\
760  & P/BO-OX   & [1,\,2]  & 765  & P/BO-OX   & [1,\,2] &   -  &   -    &  -     \\
772  & BO-OX   & [1,\,2]  & -    &        &       &   -  &   -    &  -     \\
-    & -      & -      & 980  & CO$_3$ & [3]   &   -  &   -    &  -     \\
1073 & P/BO-OX   & [1,\,2]  & 1073 & P/BO-OX   & [1,\,2] & 1073 & P/BO-OX  & [1,\,2]  \\
1192 &  P     & [1]    & 1198 & P      & [1]   &   -  &   -    &  -     \\
1241 & BO-OX     & [2]    & 1237 & BO-OX     & [2]   & 1235 & BO-OX     & [2]    \\
1260 & P      & [1]    & 1260 & P      & [1]   &  -   &    -   & -      \\
1320 & ?      & -      & -    & -      & -     & -    &   -    & -      \\
-    & -      & -      & 1398 & BA    & [1]   &  -   &  -     & -      \\
-    & -      & -      & 1457 & BA    & [1]   &  -   &  -     & -      \\
1501 & P      & [1]    & 1502 & P      & [1]   & 1505 & P      & [1]    \\
1559 & ?      & -      & 1559 & ?      & -     & 1559  & ?      & -      \\
1596 & P      & [1]    & 1597 & P      & [1]   & -    &   -    & -      \\
1606 & P      & [1]    & 1607 & P      & [1]   & 1604 &   P    & [1]    \\
1633 & ?      & -      & 1636 & ?      & -     & -    &   -    & -      \\
1665$^\dagger$ & ?      & -      & 1668$^\dagger$ & ?      & -     & -    &   -    & -      \\
-    & -      & -      & 1775 & HOCO   & [4]   & -    &   -    & -      \\
-    & -      & -      & 1885 & CO$_3$ & [3]   & -    &   -    & -      \\
-    & -      & -      & 1942 & CO$_4$ & [5]   & -    &   -    & -      \\
-    &  -     & -      & 2042 & CO$_3$ & [3]   & -    &   -    & -      \\
2138 & R-CCO  & [1]    & -    & -      & -     & -    &   -    & -      \\
-    &&& -    & -      & -    & 2845   &        & -      \\
\hline
\enddata
\tablecomments{ P\,=\,Phenol, BO-OX\,=\,Benzene oxide/oxepine, BA\,=\,Benzaldehyde (tentative assignment), ?\,=\,Unassigned bands, $\dagger$\,=\, maximum of a broad band spanning 1650-1700\,\cmu}
[1] This work, \citet{Piacentino2024ApJ...974..313P}, and \citet{parker1999photochemical}, [2] \citet{Schleif2020AngCh} [3] \citet{Moll65}, [4] \citet{Ioppolo2011MNRAS.410.1089I}, [5] \citet{Bennett2013AnaCh..85.5659B}.
\end{deluxetable*}

\subsection{Formation and destruction rates in the fiducial experiments}\label{sub:growth curves}

\begin{figure}[thb]
\centering
\includegraphics[width=\columnwidth]{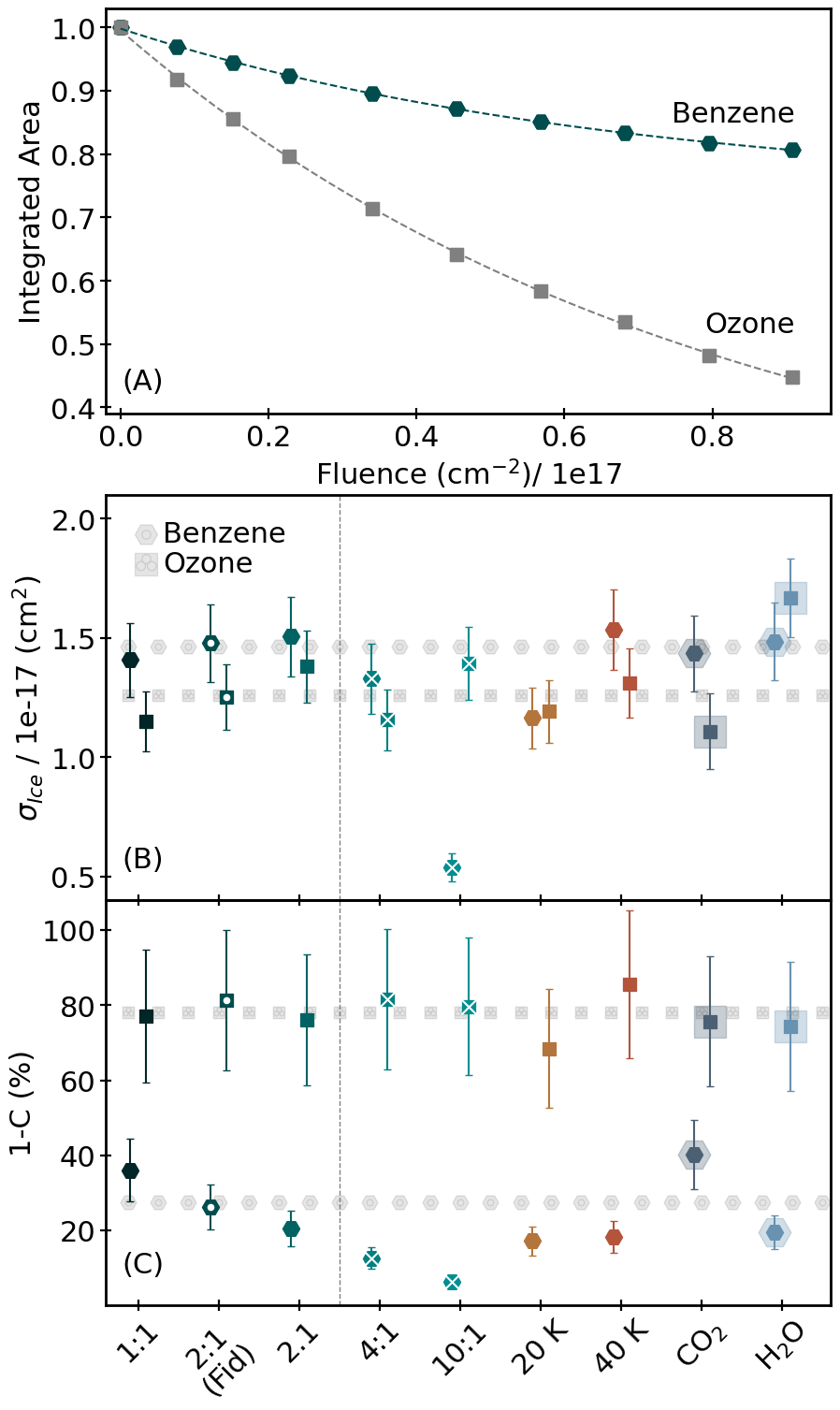}
\caption{ (A) Decay curves of benzene and ozone during ice irradiation, shown as a function of photon fluence. (B) Destruction cross sections of benzene (hexagons) and ozone (squares) under different experimental conditions. Error bars represent 11$\%$. (C) Percentage of benzene and ozone consumed in each experimental scenario. Error bars represent 23 $\%$. 
In panels B and C, the horizontal lines indicate the average of the first three data points, which are used to determine the uncertainties. The fiducial experiment (Fid) is highlighted with a white dot, while data points marked with a white x correspond to the ratio variation experiments. For easier visual distinction, the data points of the \COO and H$_2$O experiments are shown with a halo.}
\label{fig: reactant destruction}
\end{figure}

\begin{figure} [h!]
\centering
\includegraphics[width=\columnwidth]{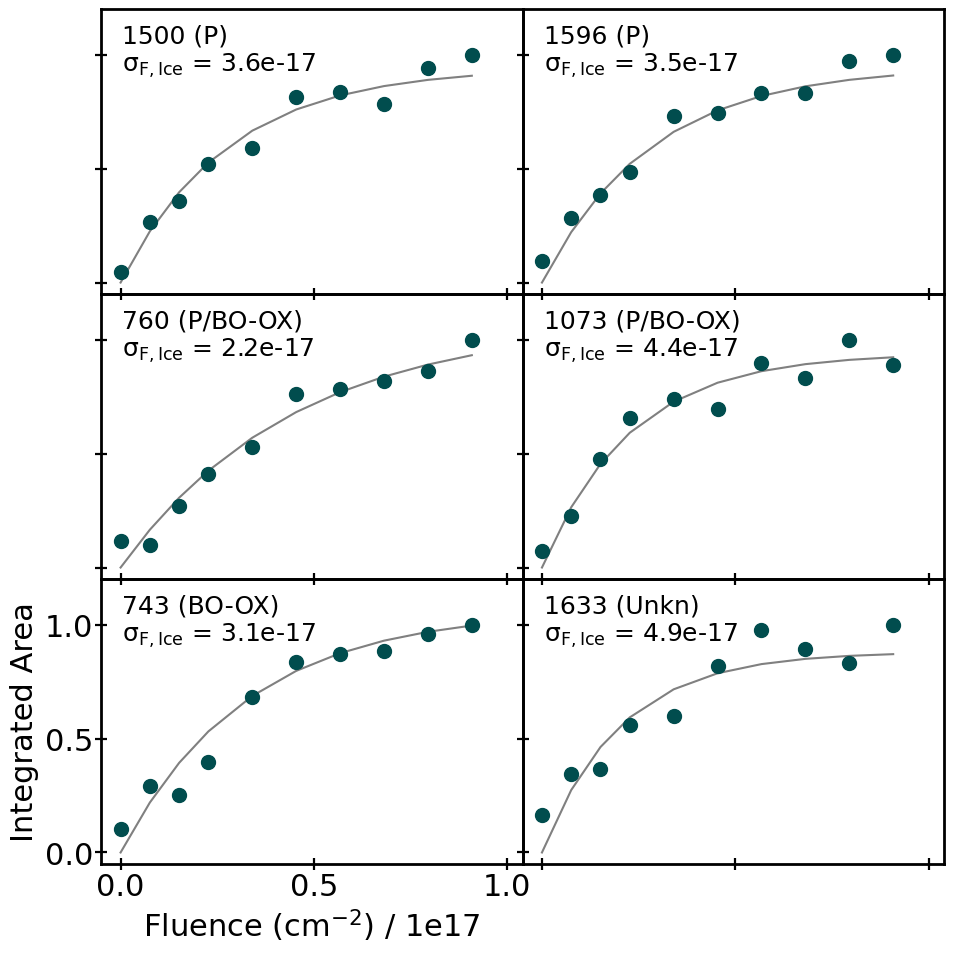}
\caption{Growth curves and formation cross sections, $\sigma$$_{F,Ice}$ (cm$^2$), of the main product bands formed in the fiducial 10 K experiment. The uncertainty on the cross sections  is of 23\,$\%$ on average, however it varies for each band (see, \S\ref{sec:uncertainty}).}
\label{fig:fit fiducial}
\end{figure}

We next evaluate the destruction kinetics of the reactants (Fig. \ref{fig: reactant destruction}) and the formation cross sections of the main product bands (Fig. \ref{fig:fit fiducial} and Fig. \ref{fig:parameters}) as a function of fluence, in order to study the reaction kinetics.
The steady-state yield of the products are reported in units of IR area (Fig. \ref{fig:correlation}) and not in ML because the IR band strengths of these molecules are either not available or have high uncertainties. We use the destruction of benzene as a proxy for the formation of C$_6$H$_6$O isomers, under the assumption that the loss of the reactant occurs exclusively via reactions with oxygen atoms.

In the fiducial experiment, by the end of photon exposure 81$\%$ of the initial \OOO and 26$\%$ of the initial benzene are consumed, respectively. The cross sections are similar, close to 10$^{-17}$ cm$^2$ for benzene and ozone (Fig. \ref{fig: reactant destruction}).
The corresponding growth curves for the main molecular products resulting from the fiducial experiment are shown in Fig.\,\ref{fig:fit fiducial}.
Formation rates were determined for two bands associated with phenol (1501 and 1596 cm$^{-1}$), the band attributed to the benzene oxide/oxepine mixture at 743 cm$^{-1}$, the ambiguously assigned bands at 1073 and 760 cm$^{-1}$, and the unassigned band at 1633 cm$^{-1}$.
The growth curves of the spectral features at 2138 and 1559 cm$^{-1}$ are shown in Appendix \ref{App: secondary}.
Within a 23$\%$ uncertainty in the $\sigma$$_{Form}$ values, we observe similar formation rates for the 1501 and 1596 cm$^{-1}$ bands, as well as between the 743 and 760 cm$^{-1}$ bands. 
Of the ambiguously assigned bands, 760\,cm$^{-1}$ shows the lowest formation rate, while 1073\,cm$^{-1}$ and 1633\,cm$^{-1}$ band have the highest formation rate.
In contrast, the growth of the unidentified band at 1559 cm$^{-1}$ and the CO band at 2138 cm$^{-1}$ (Appendix \ref{App: secondary}, Fig. \ref{figapp: secondary product fit}) is delayed relative to other products, indicating that secondary formation mechanisms are at play.

\subsection{ Effects of Ice ratio}\label{res:ratio}

We explore whether the reaction efficiency depends on small variations in the relative ratio of benzene to ozone in the ices. We studied benzene:\OOO ice ratios of 1:1, 4:1, and 10:1, in addition to two repetitions of the 2:1 fiducial experiment (Table \ref{tab_explist}).  Due to experimental constraints on the ozone flow, we were not able to investigate ice mixtures in which ozone is in excess.

\begin{figure} [t]
\centering
\includegraphics[width=\columnwidth]{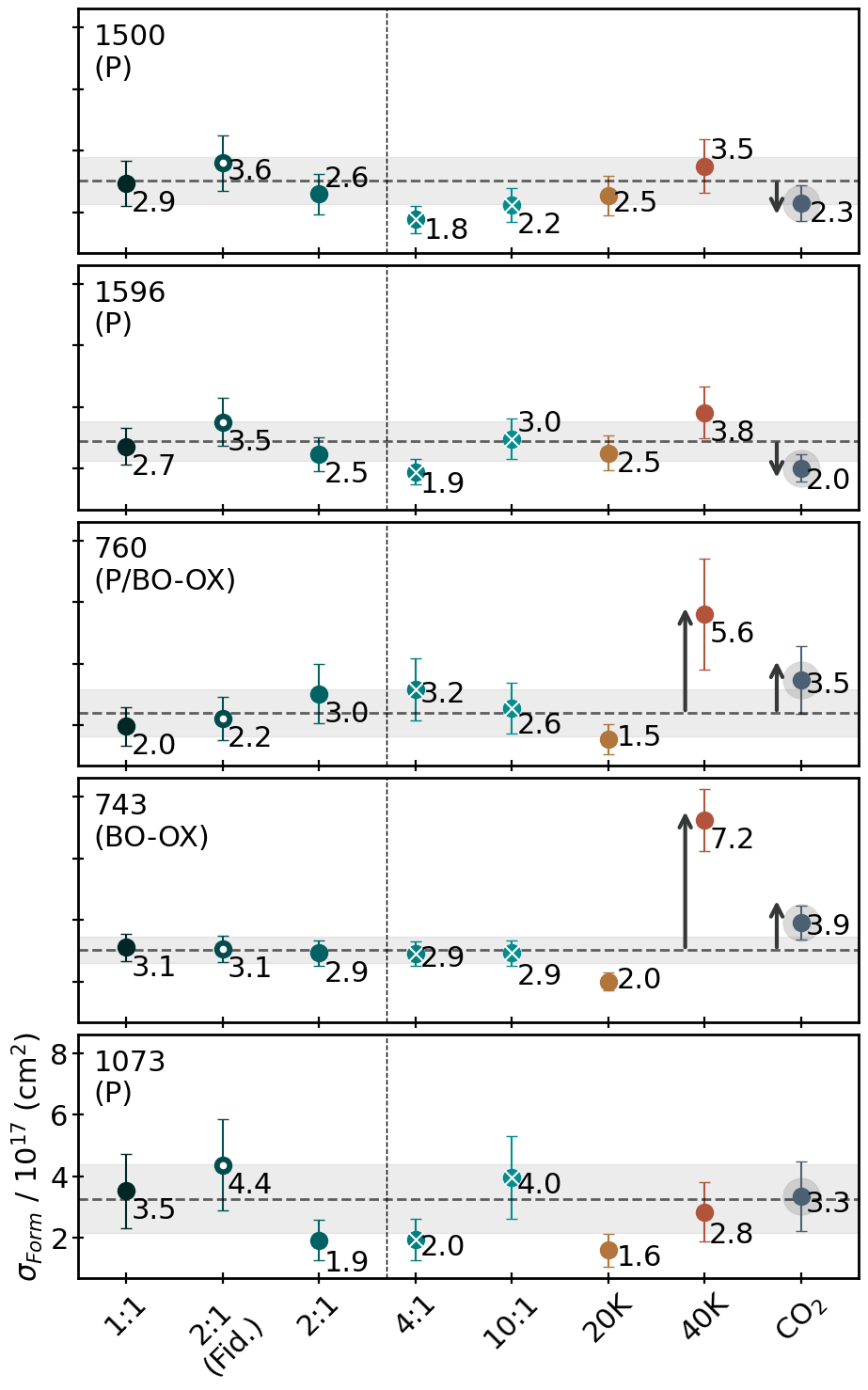}
\caption{Variation of the formation cross sections of the main product bands with temperature, ratio, and ice mixture, shaded region correspond to the uncertainty on the average value (dotted line) of the first three experiments including the fiducial (Fid.). For details on the uncertainties magnitude see \S\ref{sec:uncertainty} and App. \ref{app:uncertainty rep}. Arrows in the figure indicate specific features referenced in the text. For easier visual distinction, the fiducial experiment (Fid) is highlighted with a white dot, while data points marked with a white x correspond to the ratio variation experiments. Finally, the data point of the \COO experiment is shown with a halo.}
\label{fig:parameters}
\end{figure}

We find that at ice ratios of or below 2:1, there is no significant difference in the destruction cross sections 
of benzene and ozone (Fig. \ref{fig: reactant destruction}). In this regime, the product formation cross sections also vary only slightly (Fig. \ref{fig:parameters}), remaining within measurement uncertainty. At higher benzene concentrations, the destruction cross sections and the steady-state yields of ozone remain essentially unchanged. However, a trend emerges for benzene where, due to reduced availability of oxygen atoms, both the destruction rate and the extent of depletion decrease as its relative abundance in the mixture increases. Despite these variations in the steady-state yield of benzene, the formation cross sections of the products do not change significantly as the ice ratio shifts (Fig. \ref{fig:parameters}). However, consistent with the trend observed for the steady-state yield of benzene, the yield of product molecules decreases as the mixtures become less diluted (Fig. \ref{fig:correlation}).

\subsection{ Effects of Temperature}\label{res:temp}

We investigate the effect of ice temperature on the destruction and formation cross sections by repeating the fiducial experiments at 20 and 40\,K. Higher temperatures were not explored, as ozone begins to desorb at around 50-60\,K \citep{brann2025methyl}.  

The consumption of benzene decreases slightly with increasing temperature, from 26\,$\%$ at 10\,K to 17-18\,$\%$ at higher temperatures, while the final consumption of ozone does not vary as the temperature increases. The destruction cross sections for ozone and benzene agree with the fiducial experiment within uncertainties (Fig. \ref{fig: reactant destruction}).

\begin{figure} [h!]
\centering
\includegraphics[width=\columnwidth]{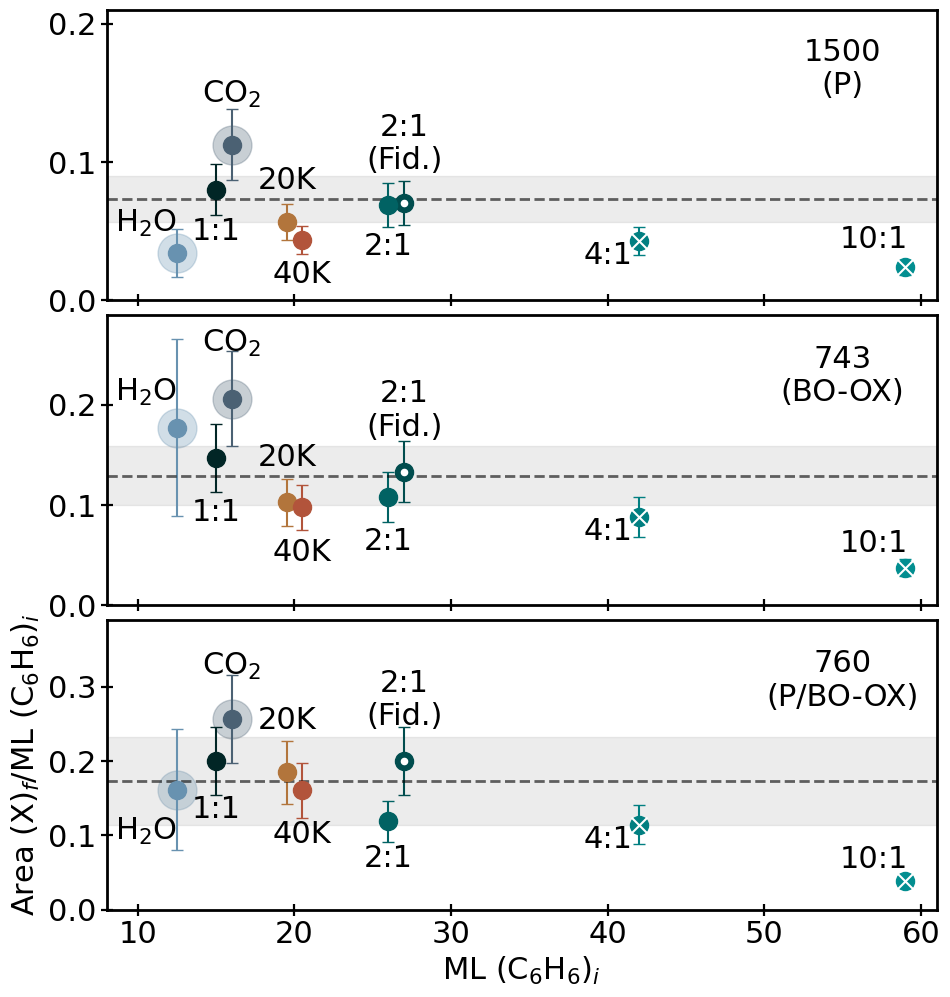}
\caption{Variation of the steady-state abundances of three main product bands relative to the initial benzene coverage (x-axis), shown across different temperatures, mixture ratios, and ice compositions. The shaded region indicates the uncertainty around the average value (dotted line). For details on uncertainty estimation, see \S\ref{sec:uncertainty} and Appendix \ref{app:uncertainty rep}.
Note that for the H$_2$O experiments, we report spectral area at the end of the experiment, as growth curves could not be constructed in this scenario. Consequently, these data points have larger uncertainties compared to others. For easier visualization, the fiducial experiment is highlighted with a white dot, while data points marked with a white “x” represent ratio variation experiments. Data from the CO$_2$ and H$_2$O ice mixtures are shown with a halo effect.}
\label{fig:correlation}
\end{figure}

As shown in Fig. \ref{fig:parameters}, most product bands also exhibit no significant change in formation cross sections with temperature. However, the band at 743 \cmu (BO/OX) and the band at 760 cm$^{-1}$ (P-BO/OX) are notable exceptions. At higher temperatures, these increase by up to a factor of three compared to the 10 K values, although their steady-state intensities remain largely unaffected (Fig. \ref{fig:correlation}). The 743\,cm$^{-1}$ band, attributed solely to benzene oxide/oxepine, shows a slightly stronger temperature dependence compared to the 760\,cm$^{-1}$ band. The latter displays an intermediate temperature dependency in the formation cross section, which is consistent with a mixed contribution from both phenol and benzene oxide/oxepine. This comparison supports the interpretation that the 760\,cm$^{-1}$ feature arises from overlapping signals of both phenol and the benzene oxide/oxepine isomers.

Overall, the results indicate a consistently efficient formation of oxygenated benzene products from the reaction with O($^1$D) in the ice, with minimal influence from the specific conditions of the surrounding ice matrix. However, temperature affects the relative reaction kinetics between phenol and benzene oxide/oxepine. While phenol formation exhibits the highest rates at 10 K, increasing the temperature enhances the rate of benzene oxide/oxepine formation relative to phenol.

\subsection{Reaction products and cross sections in diluted ice mixtures}\label{res:mixture}

\begin{figure*} [t]
\centering
\includegraphics[width=\textwidth]{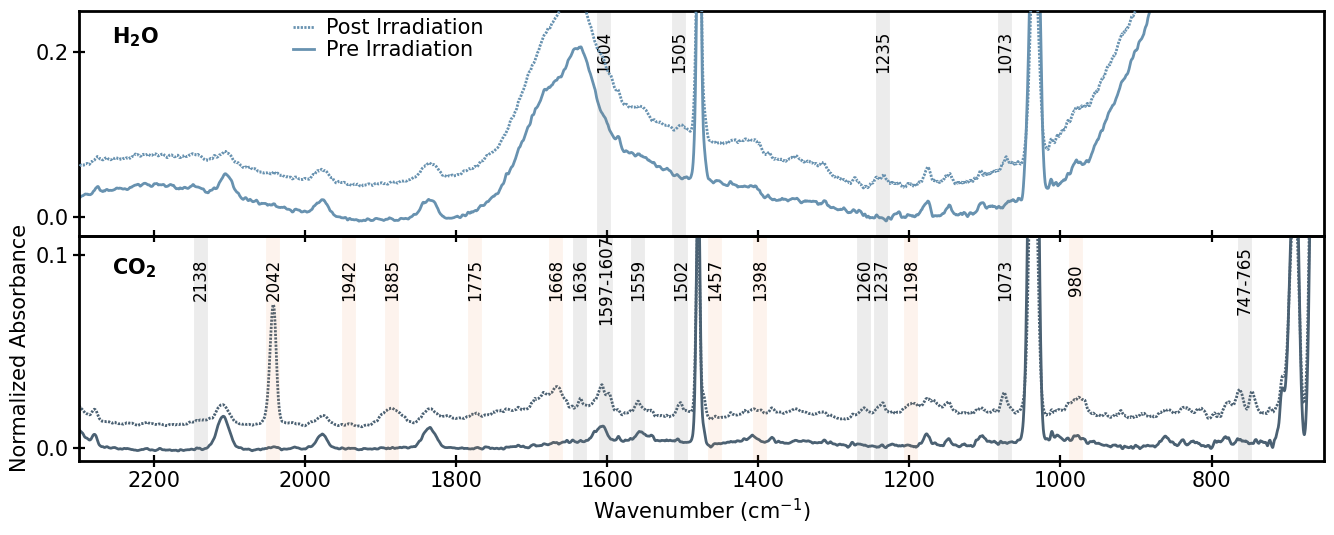}
\caption{Pre and post irradiation spectra of the H$_2$O (top panel) and \COO (bottom panel) containing ice mixtures. Product features due to \COO chemistry are highlighted in orange. While other products bands are marked in grey.}
\label{fig:mixturespectra}
\end{figure*}

The spectra of the H$_2$O and \COO ice mixture experiments are shown in Fig. \ref{fig:mixturespectra}, and the IR bands formed during the mixed ices irradiation experiments are listed in Table \ref{tab:productsoffiducial}.

In the water-containing mixture, many small spectral features are significantly muted compared to the fiducial ice due to the overalp with the strong water features. However, as shown in Fig. \ref{fig: reactant destruction}, the destruction cross sections and yields in the water-diluted experiment (1.5\,$\times$10$^{-17}$\,cm$^2$ and 20\,$\%$ for benzene and 1.7\,$\times$10$^{-17}$\,cm$^2$ and 74\,$\%$ for ozone) are not significantly different from those in the fiducial experiment, with water depletion remaining marginal ($<$\,1\,$\%$). Assuming that benzene is mainly destroyed by the reaction with O atom from O$_3$, this suggests that the presence of  water does not interfere with the efficiency of the reaction.
Furthermore, our data show the formation of key phenol bands at 1505 and 1604 $^{-1}$, along with the more ambiguous bands at 1073 and 1235 cm$^{-1}$, which may be attributed to both phenol and the benzene oxide/oxepine mixture. Additionally, the unassigned band at 1559 cm$^{-1}$ in the fiducial experiment is also observed during irradiation of the water-containing ice. Due to the muting effect of water on the IR features, we were unable to construct growth curves for this experimental condition. We note however that the yield of products in the IR spectra relative to the initial benzene available is lower for phenol in the water mixture compared to the fiducial experiment while it remains substantially unvaried for the other products (Fig. \ref{fig:correlation}). 

In the presence of CO$_2$, we observe all product bands seen in the fiducial ice, but also additional features which we attribute to products resulting from CO$_2$ and oxygen atom chemistry, including CO$_3$ at 2043, 1885, and 980 cm$^{-1}$ \citep{Moll65}, and CO$_4$ at 1942 cm$^{-1}$ \citep{Bennett2013AnaCh..85.5659B}. HOCO is also formed, indicated by a band at 1775 cm$^{-1}$ \citep{Ioppolo2011MNRAS.410.1089I}. Lastly, a feature at 1198 cm$^{-1}$ remains unassigned. In \COO ice, 40$\%$  of the benzene (twice that of the fiducial experiment) is consumed but the destruction cross section of 1.4\,$\times$10$^{-17}$\,cm$^2$ is consistent with the fiducial experiment.
The ozone destruction is slightly reduced compared to the fiducial experiments (74$\%$) and proceeds at a rate of 1.7\,$\times$10$^{-17}$\,cm$^2$ (Fig. \ref{fig: reactant destruction}). The abundance of CO$_2$ remains largely unchanged over the course of the experiment, with only about 1$\%$ destruction observed by the time steady state is reached. 
The growth curves of the bands assigned to phenol show a modest decrease in the formation cross section relative to the fiducial case (Fig. \ref{fig:parameters}), with a $\sim$25$\%$ reduction in the $\sigma$$_{Form}$ values for the 1500 and 1596 cm$^{-1}$ bands, respectively. 
In contrast, the features attributed to the benzene oxide/oxepine mixture have a slightly higher formation cross section in the presence of CO$_2$, with a $\sim$30$\%$ increase for the 743 cm$^{-1}$ band and a more substantial 60$\%$ increase for the 760 cm$^{-1}$ band compared to the fiducial case. The steady-state yields of the products in the CO$_2$ mixture are also higher, showing an increase by a factor of $\sim$\,1.5 compared to the fiducial experiment (Fig. \ref{fig:correlation}).

\section{Discussion} \label{sec:disc}

\subsection{Reaction mechanisms}
Singlet oxygen atoms are 1.98 eV higher in energy than their ground-state counterparts \citep{suijker2024unraveling}, and as a result, exhibit high reactivity. They readily undergo barrierless reactions that are independent of thermal conditions \citep{bergner2017methanol, bergner2019oxygen,nunez2018rate,lin1973reactions}, but rather depend solely on collisions with an acceptor species, benzene in our case. Ozone itself can also act as an acceptor, leading to the formation of molecular oxygen. This reaction is often the dominant quenching mechanism for O($^1$D) atoms in ozone matrices accounting for $\sim$30$\%$ of the O($^1$D) produced \citep{matsumi2003photolysis,giachardi1972photolysis}. While this process can also lead to the formation of ground state oxygen atoms, its contribution in our experiments is minimized by using a benzene dominated ice, which significantly suppresses O($^1$D) quenching by collisions with ozone and maximizes the reaction of O($^1$D) with benzene.

At 254 nm, the wavelength used for ice irradiation, benzene does not dissociate into radicals. This lack of benzene dissociation is demonstrated experimentally by two additional experiments, one run using pure benzene ice and another with benzene embedded in an inert xenon matrix, both irradiated at 254\,nm (App. \ref{app:blankbenzene}). We do not observe benzene disappearance in either of these experiments, and, similarly to our fiducial experiment, there is no evidence for carbon chain products in the IR spectra. We also do not observe the formation of bands indicative of benzene isomerization in either pure or matrix isolated ices. Therefore, we assume that electronic photoexcitation of benzene molecules, which is possible upon irradiation at 254 nm and can lead to the formation of several isomers \citep{toh2015uv}, plays only a minor role in our system. However, it remains possible that benzene isomers are formed on timescales or at concentrations that render them undetectable under our experimental conditions, and future experimental or theoretical studies are needed to resolve this. As a result, we attribute all observed chemistry to the reaction of benzene, and benzene derivatives, with oxygen atoms that form from ozone dissociation.
Based on previous studies involving aliphatic hydrocarbons, their reaction with singlet oxygen can lead to two main classes of reactions, the insertion of the oxygen atom into a C–H bond, and the insertion into, or addition to, a C–C multiple bond. This mechanism has already been observed to be successful in ices for C$_{(1-2)}$ hydrocarbons, such as in the formation of methanol and ethanol from methane and ethane \citep{bergner2017methanol, bergner2019oxygen,daniely2025photochemical}. In our case, the reaction of benzene with oxygen atoms appears to follow the same behavior, yielding both the alcohol and the C-C addition products.

The variations in the formation cross sections of the products obtained across different experimental conditions may offer clues about the chemical mechanisms at play in the ice (Fig. \ref{fig:parameters}).
Specifically, the formation rates of the bands at 1500 and 1596\,\cmu, assigned to phenol, remain unchanged as the temperature increases from 10 to 40 K, consistent with a barrierless atom insertion process into the C-H bond of benzene. This was also observed by \citet{bergner2019detection} for aliphatic hydrocarbons. Conversely in the formation of benzene oxide and oxepine, which implies interaction of the oxygen atom with a C-C bond of benzene, there is a measurable increase in the formation cross-section at 40K. This suggests that an energy barrier is involved in this O-atom addition/insertion and/or that additional formation channels open up at higher temperatures.

In particular, it has been shown that in the gas phase, quenching of O($^1$D) by O$_2$ molecules 
%, the secondary dissociation product of ozone, 
produces O($^3$P) \citep{Young1967JChPh..47.2311Y}. As the temperature of the ice increases, enhanced diffusion increases quenching of O($^1$D) by O$_2$ molecules and by other matrix constituents, consequently increasing the concentration of ground-state oxygen in the ice and opening additional reaction channels between benzene and O($^3$P). In the gas phase, reactions of benzene with O($^3$P) proceed through the formation of a benzene-oxide intermediate, which subsequently dissociates to form phenoxy + H or cyclopentadiene + CO \citep{vanuzzo2021crossed}.  In an icy environment, however, this intermediate may be stabilized, which can alter the reaction outcome and allow oxygen atoms to add to the C–C bonds of the benzene ring at various positions. Addition to adjacent carbons forms 1,2-benzene oxide, while addition to opposite carbons results in 1,4-benzene oxide \citep{boocock1961reaction}. Once formed, 1,2-benzene oxide easily isomerizes to oxepine, a process favored in condensed-phase environments due to a low activation barrier of approximately 7 kJ/mol \citep{vogel1967benzene}, promoting the formation of both benzene oxide and oxepine. The observed increase in the formation cross-sections for the 760 and 743 cm$^{-1}$ bands at 40 K can therefore be attributed to the enhanced diffusion, which allows reactions between benzene and O($^3$P). While the presence of additional pathways would be expected to increase product yields, the higher temperature also promotes more efficient desorption of the atomic species, leading to only small variations in steady-state yields for all products compared to the fiducial case (Fig. \ref{fig:correlation}).

In the presence of CO$_2$, we observe a slight decrease in the formation rates of the bands associated with phenol, and a slight increase in the bands at 743 and 760\,cm$^{-1}$, which are attributed to benzene oxide and oxepine (Fig. \ref{fig:parameters}). While these variations are close to the uncertainty limits, they may still offer some clues about underlying mechanisms. The reduction in phenol formation rate, is likely due to a decreased frequency of collisions between O($^1$D) and benzene, in favor of collisions between singlet oxygen atoms and CO$_2$ molecules in the dilution regime of our experiment. This results in quenching of singlet oxygen by CO$_2$ rather than reaction with the organic species. This hypotheses is further supported by the appearance of several \COO derived products formed through reactions with oxygen atoms. However, the steady state yield of the 1500 \cmu band appears marginally increased, which cannot be explained by the mechanism described above, suggesting that other factors may be involved and should be explored in future computational studies. The lack of decrease but rather slight enhancement, of the formation rate of  the 743 and 760 \cmu bands in the presence of \COO suggests that additional mechanisms may be contributing to the formation of benzene oxide/oxepine under these conditions. In particular, we suspect that \COO might act as secondary oxygen atoms source in the ices. CO$_2$ is not directly dissociated by 254\,nm photons \citep{nonpCruz-Diaz2014AA...562A.120C}, however, it can react with O($^1$D) to form CO$_3$. CO$_3$ is initially produced in an excited (hot) state and can subsequently relax dissociating to CO$_2$, releasing an O($^3$P) atom in the process \citep{baulch1966isotopic}. As a result, some of the O($^1$D) atoms in the ice are effectively converted to O($^3$P), which are not able to insert in the C-H bond to form phenol, but may be able to react with the C-C bond to form the benzene oxide/oxepine pair. The opening of a new reaction channel is also consistent with the slight increase in the steady state yield observed for the 743 and 760 \cmu bands (Fig. \ref{fig:correlation}).   

Finally, we are unable to draw significant conclusions about the reaction mechanisms in the water-rich ice experiments, as most product bands are heavily muted, making it impossible to extract reliable growth curves. Phenol is the only oxygenated product clearly observed in the \ce{C6H6}:\ce{O3}:\ce{H2O} system. This may be partly due to the strong IR absorption bands of water, which can obscure weaker signals from other species such as benzene oxide or oxepine. However, the absence of these features may also reflect underlying chemistry because in polar and amphoteric environments like water ice, oxepine is expected to isomerize readily to phenol in the presence of Brønsted or Lewis acids \citep{vogel1967benzene,Schleif2020AngCh}.

\subsection{Rate comparison with smaller molecules}

To contextualize our measured formation rates for phenol, we compare them with the work of \citet{bergner2019detection} on the reactivity of smaller hydrocarbons with singlet oxygen in ices. However, a direct comparison between the ethanol formation cross section reported by \citet{bergner2019detection} and our derived formation cross section for phenol is not straightforward due to differences in experimental procedure. The most relevant difference is the use of \COO as the oxygen source in the ethane-to-ethanol experiments by \citet{bergner2019detection}, versus O$_{3}$ in our benzene-to-phenol experiments. This distinction intrinsically leads to a different number of oxygen atoms produced per incident photon, as the photodissociation cross sections of of \COO and O$_3$ molecules differ. To enable a more meaningful comparison, we instead plot the formation data as a function of the number of O(\textsuperscript{1}D) atoms produced in the ice. The data for ethane are reproduced from \citet{bergner2019detection} with permission. 

\begin{figure} [h!]
\centering
\includegraphics[width=\columnwidth]{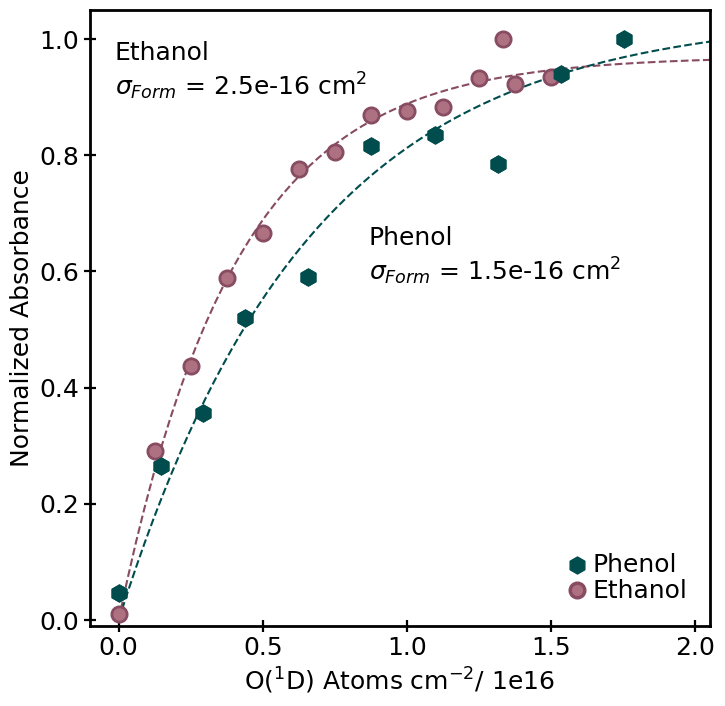}
\caption{Growth curves as a function of O($^1$D) exposure for ethanol (purple circles) and phenol (teal hexagons) ices, formed from ethane and benzene ices, respectively. Dashed lines represent fits to the growth model in Eq. \ref{eq. integrated law}, with the corresponding formation cross sections indicated.}
\label{fig:ethane}
\end{figure}

For the ethane\,+\,\COO experiment, we estimated the number of O($^{1}$D) atoms produced assuming a quantum yield of unity for the \COO to O($^{1}$D) process, and using the photodissociation cross section for \COO, derived from the experimental data in \citet{bergner2019detection}, of 2.5\,$\times$\,10$^{-19}$ cm$^2$.  
To determine the O($^{1}$D) availability in the benzene\,+\,O$_{3}$ experiment, we used the cross section listed in Table \ref{tab_explist} from our fiducial experiment, again assuming a quantum yield of one for O($^{1}$D) formation \citep{matsumi2003photolysis}. The resulting comparison, shown in Figure \ref{fig:ethane}, indicates that the formation cross section for phenol from benzene, as a function of O($^{1}$D) atoms available in the ice, is comparable in magnitude to that obtained for ethanol in the ethane experiment.
The similarity in formation cross sections for phenol and ethanol indicates that both reactions proceed via efficient barrierless O($^1$D) atom insertion into C–H bonds. The comparable reactivity suggests that neither the number of available insertion sites nor the presence of multiple bonds significantly affects the reaction efficiency, confirming a generally structure-insensitive mechanism for O($^1$D) insertion into hydrocarbons in the solid phase.

However, by the same rationale, smaller $\pi$-systems such as ethene (C$_2$H$_4$) and acetylene (C$_2$H$_2$) would also be expected to exhibit similar reactivity \citep{alkorta2023nucleophilicities}. Indeed, theoretical studies predict that O($^1$D) insertion into ethene is energetically favorable \citep{daniely2025photochemical}. Nonetheless, the resulting products have not been observed in ice experiments \citep{bergner2019oxygen}. This difference may arise from the lower stability and increased susceptibility to dissociation of the alcohol products formed in C$_2$H$_4$ and C$_2$H$_2$ ices \citep{daniely2025photochemical}. 

\section{Astrochemical implications} \label{astrochem}

\subsection{Estimating phenol formation in astrophysical ices} \label{icemodel}

Our experimental results suggest that oxygen insertion reactions offer a viable ice-phase pathway for forming oxygenated aromatic compounds such as phenol and benzene oxide/oxepine. These low-temperature reactions may contribute to molecular complexity during the early stages of star and planet formation, when simple molecules are frozen into interstellar or protostellar ices. However, the experimental conditions used do not directly replicate the much more dilute environments of realistic ices. In particular, the local availability of reactive singlet oxygen atoms is significantly lower in space, and CO$_2$, an abundant component of interstellar ices, is likely the primary source of O($^1$D) in such environments \citep{Pontoppidan08}. To account for this, we developed a simplified model to estimate the formation of C$_6$H$_6$O species from benzene embedded in realistic astrophysical ices, where CO$_2$ acts as the primary source of O($^1$D) through photodissociation \citep{Okabe78}. 

We focused on the local generation of O($^1$D) near a benzene molecule via photodissociation of adjacent CO$_2$ molecules. To explore this, we considered two astrophysically relevant scenarios. The first is a high-concentration case representative of low-mass young stellar objects (YSOs), where the CO$_2$/H$_2$O ratio is approximately 28$\%$ \citep{boogert2015observations}, with about two-thirds of the CO$_2$ mixed into H$_2$O-rich ices \citep{Pontoppidan08}. The second is a low-concentration case based on observations of comet 67P, where the CO$_2$/H$_2$O ratio is 7.5$\%$, but only around 13$\%$ of the CO$_2$ is mixed with water ice resulting in about 1:100 mixing ratio \citep{rubin2023volatiles}. 

Assuming that benzene is embedded in a water-rich ice, we first estimate the number of CO$_2$ molecules that may be adjacent to a benzene molecule in both the high and low concentration scenario. Based on geometric considerations, approximately 17 water molecules can fit within the first solvation shell of benzene; however, gas-phase simulations have shown that up to 31 water molecules can arrange around a benzene molecule \citep{choudhary2015spatial}. For our model, we consider both values to account for the possible range of water molecules surrounding benzene. Applying the CO$_2$/H$_2$O ratios discusses above to these ranges yields approximately 3–6 CO$_2$ molecules in the high concentration astrophysical scenario, and $<$ 1 CO$_2$ molecule in the low concentration scenario.

The photodissociation of these neighboring CO$_2$ molecules by UV photons can be modeled as a first-order kinetic process, producing excited-state oxygen atoms O($^1$D),

\begin{equation}
\rm D_{\mathrm{(CO_2)}}(t) = 1 - e^{(-\sigma_{\mathrm{CO_2}} \cdot \phi \cdot t)}
\end{equation}

where $\phi$ $\simeq$10$^4$ cm$^{-2}$ s$^{-1}$ is the flux of dissociating radiation in interstellar clouds shielded from interstellar UV \citep{Shen04}, and $\sigma_{\mathrm{CO_2}} = 2.4 \times 10^{-18}~\mathrm{cm}^2$ is the experimental cross section measured on the same experimental set-up (Narayanan et al. 2025, in prep.). We note that this cross section value is in reasonable agreement with previous values from the literature \citep{rafa2015uv,gerakines1996ultraviolet}.  

The resulting number of O($^1$D) atoms under the assumption of a reaction efficiency of unity \citep{Okabe78} is,

\begin{equation}
\rm F_{O(^1D)}(t) = D_{\mathrm{(CO_2)}}(t) \cdot N_{(CO_2)}
\end{equation}

where $D_{\mathrm{(CO_2)}}$(t) is the fraction of dissociated CO$_2$, and $N_{(CO_2)}$ is the number of initial CO$_2$ molecules. 

\begin{figure}[thb]
\centering
\includegraphics[width=1\columnwidth]{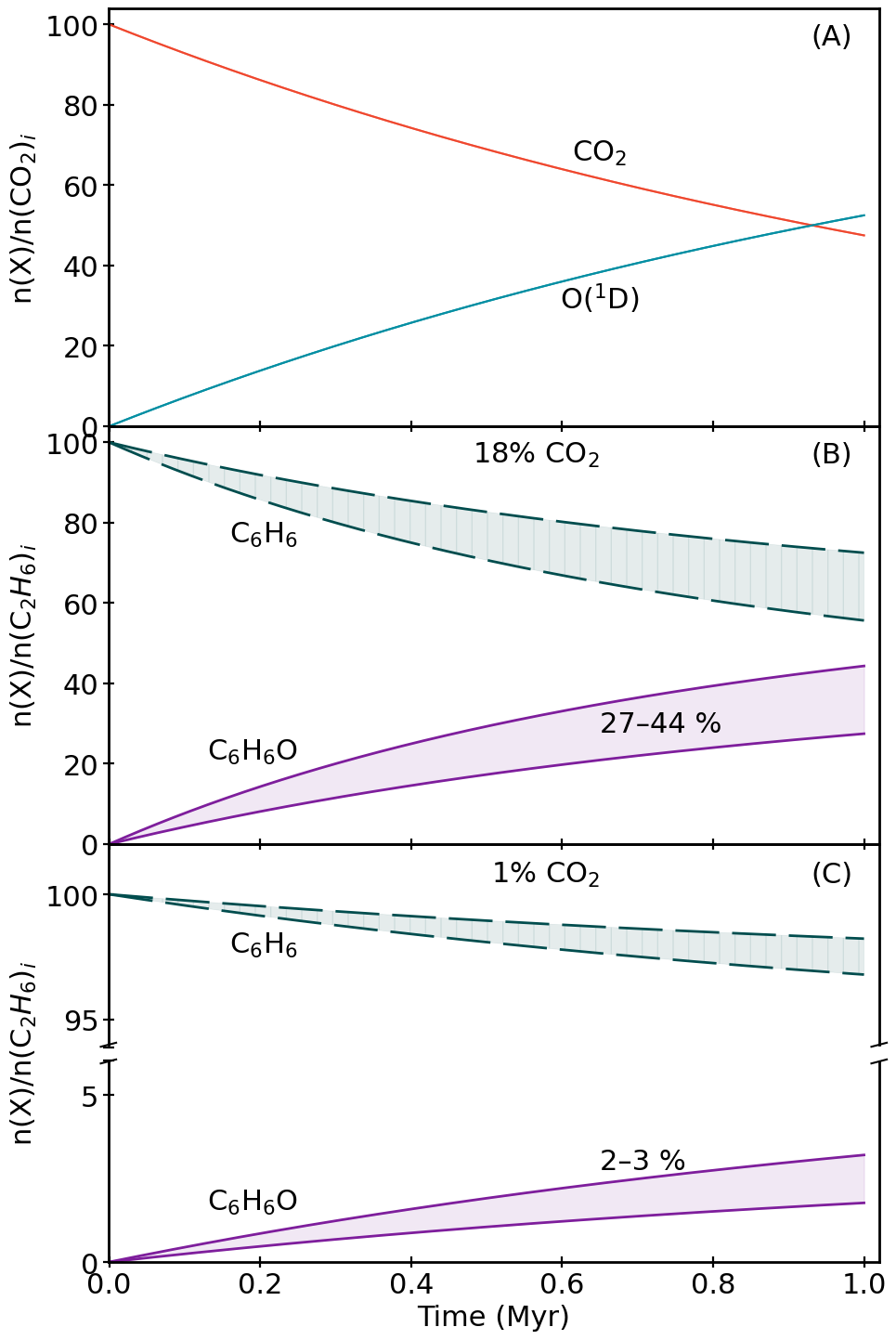}
\caption{Cumulative evolution of CO$_2$ dissociation (red) and O($^1$D) formation (blue) in ices exposed to a constant photon flux of $\phi$ $\simeq$10$^4$ cm$^{-2}$ s$^{-1}$ is the flux of dissociating radiation over 1 Myr (A). over 1 Myr (A). Corresponding percentage of benzene converted into C$_6$H$_6$O (purple) and remaining unreacted (teal) for CO$_2$:H$_2$O ice ratios of 18\,$\%$ (B) and 1\,$\%$ (C), representing high and low concentration scenarios, respectively. Shaded regions reflect the range of outcomes due to different benzene solvation shell sizes (see text for details).
}
\label{fig:co2model}
\end{figure}

We also assume that all O($^1$D) atoms generated in the immediate vicinity of benzene collide with it before being quenched by the surrounding ice matrix. Based on our experimental results (Table~\ref{tab_explist}), about 20$\%$ (Col$_{(ef)}$) of these collisions result in efficient reactions that convert benzene to C$_6$H$_6$O. Under these conditions, we further assume that all disappearing benzene is converted exclusively to C$_6$H$_6$O, with no significant competing reactions. The fraction of benzene converted into oxygenated products at any time is therefore given by

\begin{equation}
\rm F_{C_6H_6O}(t) = 1- e^{-Col_{(ef)} \cdot F_{O(^1D)}(t))}
\end{equation}

The results of this simplified model, shown in Fig. \ref{fig:co2model}, indicate that under interstellar medium (ISM) conditions where CO$_2$ is relatively abundant, between 27$\%$ and 44$\%$ of benzene can be converted into oxygenated products over the course of one million years. In contrast, when the concentration of CO$_2$ is less abundant and thus fewer reactive oxygen atoms are locally available, the fraction of benzene converted is significantly lower, amounting to only about 2–3$\%$ over the same timescale. These results suggest that depending on the specific availability of singlet oxygen atoms, oxygen insertion reactions could play a major role in altering the chemical nature of aromatic molecules in realistic interstellar ices, potentially contributing to the early formation of more complex organic molecules. 

\subsection{Constraints from Astronomical and Cometary Observations}

Validating our model remains challenging due to limited observational constraints on benzene. To date, detections include the protoplanetary nebula CRL 618, where benzene has been observed in the gas phase with a column density up to $5 \times 10^{15}\,\mathrm{cm}^{-2}$ \citep{cernicharo2001infrared}. Similarly, phenol has only been tentatively detected in Orion KL \citep{kolesnikova2013millimeter}, with a reported column density of (8 ± 4)  10$^{14}$ cm$^2$ \citep{kolesnikova2013millimeter}. 
In addition, related species such as o-benzyne (o-C$_6$H$_4$), phenylacetylene (C$_6$H$_5$CCH), and benzonitrile (C$_6$H$_5$CN) have been detected in TMC-1 at column densities of $\sim 5 \times 10^{11}\,\mathrm{cm}^{-2}$ \citep{cernicharo2021discovery}, $\sim 3 \times 10^{12}\,\mathrm{cm}^{-2}$ \citep{loru2023detection},  and $\sim 4 \times 10^{11}\,\mathrm{cm}^{-2}$ \citep{mcguire2018Sci...359..202M}, respectively.
% In addition, related species such as o-benzyne (o-C$_6$H$_4$) and benzonitrile (C$_6$H$_5$CN) have been detected in TMC-1 at column densities of $\sim 5 \times 10^{11}\,\mathrm{cm}^{-2}$ \citep{cernicharo2021discovery} and $\sim 4 \times 10^{-11}\,\mathrm{cm}^{-2}$ \citep{mcguire2018Sci...359..202M}, respectively. 
These detections suggest that benzene accumulates sufficiently in the ISM to undergo chemical functionalization, giving rise to a range of substituted aromatic compounds. While not fully constrained, these gas-phase column densities, combined with the low volatility of aromatic molecules \citep{Piacentino2024ApJ...974..313P}, suggest that the ice is likely enriched in aromatic and oxygenated aromatic molecules. Astrochemical models are needed to evaluate benzene abundance and reactivity in the ice, but given the sensitivity of results to the chemical networks, an accurate definition of the aromatic chemistry in both gas and ice phases is crucial.

While direct observational evidence for oxygenated aromatic compounds in interstellar environments remains limited, cometary studies provide a unique opportunity to assess whether molecules like benzene and phenol persist under solar system conditions. Aromatic molecules have indeed been detected in solar system bodies, with the European Space Agency’s Rosetta mission to comet 67P/Churyumov–Gerasimenko providing the first \textit{in situ} detections of benzene \citep{schuhmann2019aliphatic} and, tentatively, phenol \citep{hanni2023oxygen} within the comet’s coma. The observation of these molecules was made using the Double Focusing Mass Spectrometer (DFMS; \citet{balsiger2007rosina}), part of the ROSINA instrument suite onboard the Rosetta orbiter, which continuously monitored the volatile composition of the coma as the comet traveled along its orbit around the Sun. Benzene was initially identified in May 2015, shortly after the comet’s inbound equinox \citep{schuhmann2019aliphatic}. However, a more comprehensive dataset for complex organic molecules was obtained during the peak gas and dust activity shortly before the comet’s perihelion in August 2015 \citep{hanni2022identification}. During this period, the data revealed enhanced abundances of benzene relative to common cometary volatiles such as water and methanol, and also demonstrated the presence of numerous heteroatom-bearing organic species, including C$_6$H$_6$O  \citep{hanni2022identification,hanni2023oxygen,Hanni2025nitrogenA&A...699A.135H}. 
Such cometary measurements provide a valuable reference point that allows us to compare and evaluate the relevance of our laboratory-derived benzene conversion efficiencies under realistic astrophysical conditions.

While the identification of benzene in the August 2015 is considered robust (\citet{hanni2022identification}, supplementary material), the assignment of the observed C$_6$H$_6$O \citep{hanni2023oxygen} parent ion remains ambiguous, mainly due to the lack of comprehensive reference fragmentation data for all C$_6$H$_6$O isomers (NIST). In particular, isomers such as benzene oxide and oxepine lack available reference mass spectra, leaving their potential contribution to the DFMS signal unconstrained. Nevertheless, \citet{hanni2023oxygen} proposed phenol as a likely candidate based on the presence of strong diagnostic fragments and the molecule’s structural simplicity. However, in comparing the cometary observations to our experimental model (Fig. \ref{fig:co2model}), precise differentiation between isomers is not necessary, as our model focuses on C$_6$H$_6$O formation as a whole, rather than specific isomers formed in the ice.

Using data from 3 August 2015, we performed a new renormalization of previously measured fragment sums for C$_6$H$_6$O and C$_6$H$_6$ \citep{hanni2022identification,hanni2023oxygen} to derive an estimate of the C$_6$H$_6$O:C$_6$H$_6$ abundance ratio in 67P’s dusty coma. The resulting estimate yields a C$_6$H$_6$O:C$_6$H$_6$ ratio of approximately 0.2 (20$\%$).
This ratio should be treated as an upper limit, 
with an uncertainty of about 30$\%$ mainly due to overlapping phenol-related fragments, while the benzene measurement is more precise, with less than 10$\%$ uncertainty.  It is also important to note that although C$_6$H$_6$ and C$_6$H$_6$O have similar chemical structures, which suggests comparable instrument sensitivities, sensitivity factors such as ionization cross section and instrument transmission have not been experimentally determined for these molecules and need to be considered for a comprehensive quantitative results. A detail description of the uncertainty estimation can be found in \citep{hanni2022identification,hanni2023oxygen,Hanni2025nitrogenA&A...699A.135H}. Finally, we note that our estimation does not represent the comet’s bulk ices (see, e.g., \citet{rubin2019elemental,schuhmann2019aliphatic}), but rather reflects conditions in the dust-dominated coma, where typical ice species are relatively depleted compared to organics and nominal coma abundances. Consequently, estimating the C$_6$H$_6$O:C$_6$H$_6$ ratio relative to water is not possible. Nonetheless, this aligns with the expectation that some aromatic species persist on icy grains after water sublimation due to their low volatility \citep{Piacentino2024ApJ...974..313P}.

When compared to our laboratory-derived benzene conversion efficiencies, the cometary upper-limit estimate of 20±6$\%$ C$_6$H$_6$O:C$_6$H$_6$ ratio fits well within the range of 2–44$\%$ obtained from our simplified laboratory based model (Section \ref{icemodel}). While the model ignores many important ice chemical processes, the comparison still demonstrates that O insertion is one plausible explanation for the observed cometary C$_6$H$_6$O:C$_6$H$_6$ ratio. This result suggests that excited state oxygen insertion and addition chemistry is an important process in the evolution of aromatic molecules in astrophysical ices, contributing significantly to the inventory of oxygenated organics observed in comets.

\section{Conclusions}

We have explored the chemistry of benzene and singlet oxygen atoms in the ice phase. Ozone was used as an \textit{in situ} precursor for the formation of O($^1$D) upon ice exposure to 254\,nm radiation. We also explored the same reaction at different temperatures and in ices mixed with H$_2$O or CO$_2$. We found that:

\begin{itemize}
    \item Oxygen insertion in aromatic C-H bonds is a viable pathway for the formation of aromatic alcohols. The rate of formation of phenol is similar to what literature reports for smaller aromatics.
    \item Benzene oxide and oxepine are also products of the studied reaction, forming at rates comparable to phenol. The temperature dependence of their formation cross sections suggests that additional reaction pathways may occur alongside the reaction with singlet oxygen atoms.
    \item CO$_2$ enhances the reaction rate for atom addition products, likely by efficiently quenching O($^1$D) to form O($^3$P), which likely opens additional reaction channels.
    \item In water-rich ices, phenol formation is observed, but its formation rate as well as the formation of benzene oxide could not be determined due to spectral overlap with water ice bands.
    \item Overall, our experiments suggest that oxygen insertion reactions could contribute significantly to the formation of oxygenated aromatic compounds in interstellar ices, with benzene conversion yields ranging from approximately 2 to 44$\%$, depending on the availability of oxygen-forming precursors in the ice.
    \item This experimental range is consistent with cometary observations from Rosetta, where a phenol-to-benzene upper-limit ratio of about 20$\%$ was estimated in the coma of comet 67P. This agreement supports the plausibility of oxygen insertion chemistry as a pathway for aromatic functionalization under astrophysical conditions.
\end{itemize}

All the experimental data are available at 10.5281/zenodo.17216222 and 10.5281/zenodo.17216256.

This work was supported by a grant from the Simons Foundation 686302, KÖ., KÖ. E.L.P. thanks Prof. Tim Schleif, Dr. Melania Prado Merini, and Prof. Wolfram Sander for generously sharing the raw spectral data files for oxepine and benzene oxide. N.H. acknowledges the help of Rosetta/DFMS principal investigator Prof. Kathrin Altwegg
(University of Bern) who advised the reduction of the 3 August 2015 dataset archived and publicly accessible on the European Space Agency’s Planetary Science Archive. Work by
N.H. was funded by the Canton of Bern and the Swiss National Science Foundation (200020$\_$207312). A. M. acknowledges the support of the Natural Sciences and Engineering Research Council of Canada (NSERC), [funding reference number 587448]. Cette recherche a été financée par le Conseil de recherches en sciences naturelles et en génie du Canada (CRSNG), [numéro de référence 587448].

% \newpage
\bibliography{sample631}{}
\bibliographystyle{aasjournal}

\newpage
\appendix{}\label{sec:appendix}

\section{Second generation products}\label{App: secondary}

\begin{figure}[h]
\centering
\includegraphics[width=0.6\textwidth]{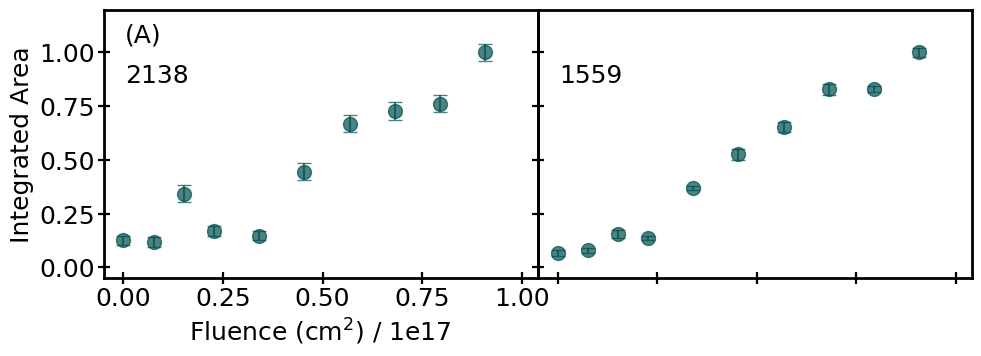}
\centering
\includegraphics[width=0.6\textwidth]{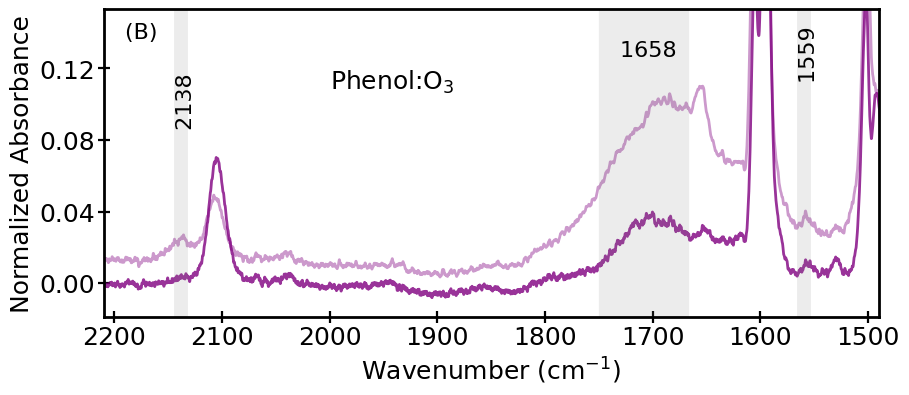}
\caption{(A) Growth curves of the 1559 and 2138 cm$^{-1}$ bands observed in the fiducial experiment.
(B) Infrared spectra of a phenol:O$_3$ ice mixture in a 1:2.5 ratio, recorded before and after irradiation with 254 nm photons at a maximum fluence of 9.2$\times$10$^{17}$ cm$^{-2}$. }
\label{figapp: secondary product fit}
\end{figure}

Figure \ref{figapp: secondary product fit}, panel A, presents the growth curves of the 1559 and 2138 cm$^{-1}$ bands observed in the fiducial experiment. These features are attributed to second-generation chemistry based on their delayed growth onset. To explore this further, we irradiated a phenol:O$_3$ ice mixture in a 1:2.5 ratio using 254 nm photons and reached a maximum fluence of 9.2$\times$10$^{17}$ cm$^{-2}$ (Figure \ref{figapp: secondary product fit}, panel B). The resulting spectra show two new features. One is a broad band centered near 1658 cm$^{-1}$, which remains unidentified. The other is a weaker shoulder appearing at 2138 cm$^{-1}$, likely arising from the formation of a ketene of general formula R–C=C=O \citep{parker1999photochemical,hudson2013ketene}. The similarities between the 1658 cm$^{-1}$ band and the signal observed in the fiducial experiment point to a common origin that may involve secondary reactivity between phenol and atomic oxygen. A feature at 1559 cm$^{-1}$ is already present in the phenol:O$_3$ ice mixture prior to irradiation, but it becomes more pronounced upon exposure to photons. This change in band intensity and shape observed during phenol irradiation likely contributes to the emergence of the similar feature in the fiducial experiment.  While more detailed experiments would be needed to confirm the assignment of these bands, we speculate that phenoxy radicals might be responsible for the band at 1559 \cmu \citep{spanget2001vibrations}, and that the broad band centered at 1658 \cmu might be due to  cyclohexadienone, which can form from isomerization of phenol \citep{vanuzzo2021crossed,zhu2003kinetics}.   

\newpage
\section{Blank experiments}\label{app:blankbenzene}

\begin{figure}[h]
\centering
\includegraphics[width=0.9\textwidth]{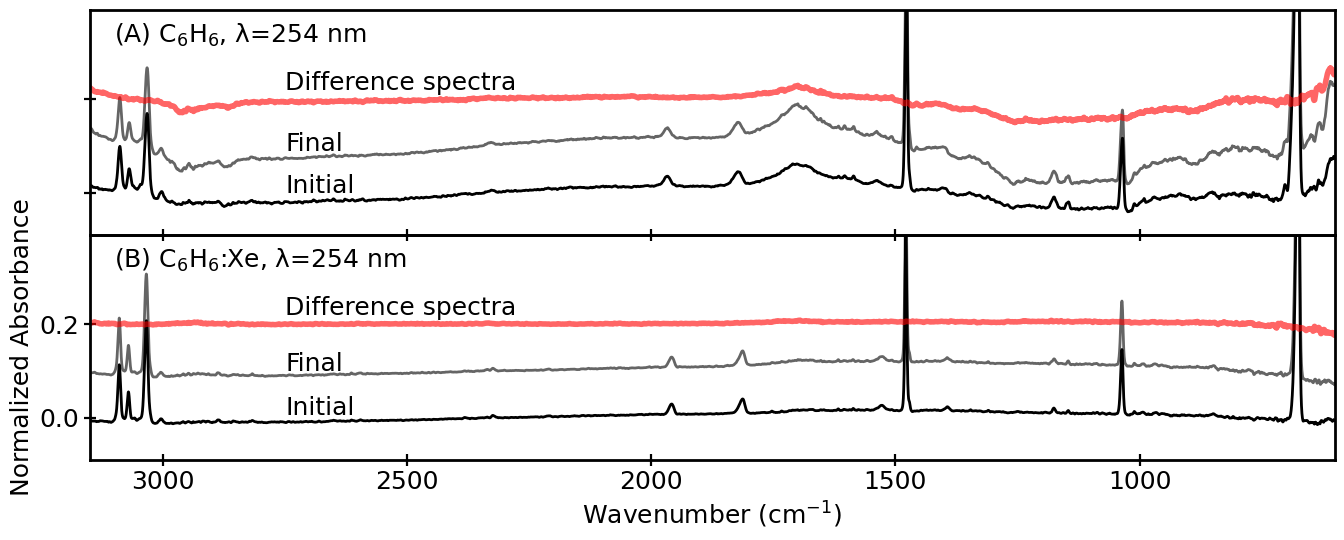}
\caption{Initial, final, and difference IR spectra for the two control experiments: (A) pure benzene ice; (B) benzene isolated in a xenon matrix.}
\label{figapp: blank}
\end{figure}

To determine whether benzene is reactive under 254 nm irradiation, we conducted two control experiments. In one of the experiment, pure benzene ice was exposed to 254 nm photons. In the other experiment, benzene was diluted in a xenon matrix and also irradiated at 254 nm. In both cases, no evidence of dissociation or product formation was observed. The difference spectra of the pure benzene ice showed no new features (Fig. \ref{figapp: blank}, panel A), and the matrix-isolated sample similarly exhibited no spectral changes (Fig. \ref{figapp: blank}, panel B). These results confirm that benzene remains photochemically inert at 254 nm under our experimental conditions.

\newpage
\section{Experiments Used for the Uncertainty Estimation}\label{app:uncertainty rep}

\begin{figure}[h]
\centering
\includegraphics[width=0.96\textwidth]{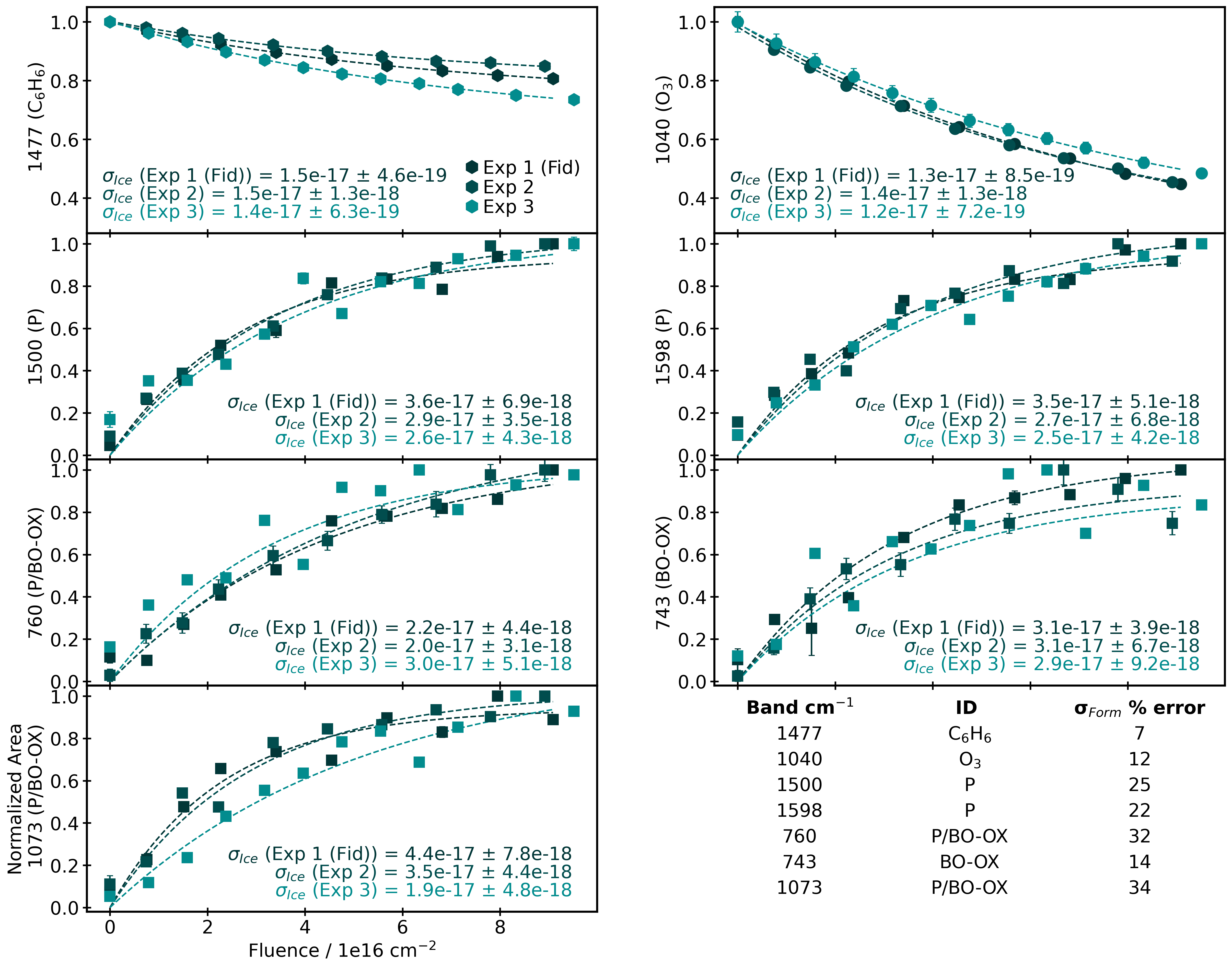}
\caption{Normalized spectral feature evolution for seven bands across three experiments (Exp 1–3), plotted as a function of fluence. Each subplot includes model fits (growth or decay) based on Eq. \ref{eq:decay} or Eq \ref{eq. integrated law}. Cross section values ($\sigma$) and their associated uncertainties are annotated for each experiment. The table in the bottom-right panel summarizes the cumulative percentage uncertainty in formation cross sections for each spectral feature.}
\label{figapp: rep}
\end{figure}

We use the spectral data from the fiducial experiment (Exp. 1 in Table \ref{tab_explist}), along with data from Experiments 2 and 3, to estimate the uncertainty in the reported formation cross sections ($\sigma_{Form}$). In Fig. \ref{figapp: rep} we show model fits (growth and decay) to the spectral features observed in all three experiments. These fits are applied to both the reactant bands (ozone and benzene) and the primary product bands. Each subplot displays the resulting decay or growth constant, along with its associated uncertainty from the fitting procedure. 
In addition to the fitting uncertainty, the total error includes the variability between experiments, estimated from the spread in fitted values, and a 5$\%$ uncertainty in fluence calibration due to the absolute lamp intensity. This calibration uncertainty is evaluated based on its effect on the resulting fitted cross sections. The cumulative uncertainties for each spectral band are summarized in the table shown in the bottom-right panel of the figure. These values are propagated into the error bars shown in \S\ref{fig:parameters} and Fig. \ref{fig: reactant destruction}.

\end{document}